\pdfoutput=1

\documentclass[11pt,twoside,a4paper,cmspaper,final,collab]{cms-tdr}
\def\svnVersion{306343}\def\cmsCernNo{2013-037}\def\cmsCernDate{\today}\def\cmsMessage{Published in Physical Review D as \href{http://dx.doi.org/10.1103/PhysRevD.92.072006}{\doi{10.1103/PhysRevD.92.072006.}}}

\begin{document}\cmsNoteHeader{SUS-14-004}

\hyphenation{had-ron-i-za-tion}
\hyphenation{cal-or-i-me-ter}
\hyphenation{de-vices}
\RCS$Revision: 306305 $
\RCS$HeadURL: svn+ssh://svn.cern.ch/reps/tdr2/papers/SUS-14-004/trunk/SUS-14-004.tex $
\RCS$Id: SUS-14-004.tex 306305 2015-10-21 12:39:04Z kiesel $
\ifthenelse{\boolean{cms@external}}{\providecommand{\cmsLeft}{top}}{\providecommand{\cmsLeft}{left}}
\ifthenelse{\boolean{cms@external}}{\providecommand{\cmsRight}{bottom}}{\providecommand{\cmsRight}{right}}
\ifthenelse{\boolean{cms@external}}{\providecommand{\CL}{C.L.\xspace}}{\providecommand{\CL}{CL\xspace}}
\newcommand{\CLs}{\ensuremath{\mathrm{CL}_\mathrm{s}}\xspace}

\newcommand{\MR}{\ensuremath{M_\mathrm{R}}\xspace}
\newcommand{\MRz}{\ensuremath{M_\mathrm{R}^0}\xspace}
\newcommand{\Rtwo}{\ensuremath{\mathrm{R}^2}\xspace}
\newcommand{\Rtwoz}{\ensuremath{\mathrm{R}^2_0}\xspace}
\newcommand{\R}{\ensuremath{\mathrm{R}}\xspace}
\newcommand{\MRT}{\ensuremath{M_\mathrm{T}^\mathrm{R}}\xspace}
\providecommand{\cPV}{\ensuremath{\cmsSymbolFace{V}}\xspace}

\newcommand{\fakeRate}{\ensuremath{ f_{\Pe\to\gamma} }\xspace}
\newcommand{\ciso}{\ensuremath{ I_\pi }\xspace}
\newcommand{\niso}{\ensuremath{ I_\mathrm{n} }\xspace}
\newcommand{\piso}{\ensuremath{ I_\gamma }\xspace}
\newcommand{\ptStar}{\ensuremath{\pt^{*}}\xspace}
\newcommand{\HTstar}{\ensuremath{\HT^{*}}\xspace}

\newcommand{\ggamma}{\ensuremath{ \gamma_\text{tight} }\xspace}
\newcommand{\fgamma}{\ensuremath{ \gamma_\text{loose} }\xspace}
\newcommand{\egamma}{\ensuremath{ \gamma_\text{pixel} }\xspace}

\newcommand{\lumiData}{19.7\fbinv}

\cmsNoteHeader{SUS-14-004}
\title{Search for supersymmetry with photons in \texorpdfstring{$\Pp\Pp$}{pp} collisions at \texorpdfstring{$\sqrt{s}=8\TeV$}{sqrt(s)=8 TeV}}

\date{\today}

\abstract{ Two searches for physics beyond the standard model
  in events containing photons are presented. The data sample used corresponds
  to an integrated luminosity of 19.7\fbinv of proton-proton collisions at
  $\sqrt{s}=8\TeV$, collected with the {CMS} experiment at the {CERN} {LHC}.  The
  analyses pursue different inclusive search strategies. One analysis
  requires at least one photon, at least two jets, and a large amount of
  transverse momentum imbalance, while the other selects events with at
  least two photons and at least one jet, and uses the razor variables to search for
  signal events. The background expected from standard model processes is evaluated mainly
  from data.  The results are interpreted in the context
  of general gauge-mediated supersymmetry, with the next-to-lightest supersymmetric
  particle either a bino- or wino-like neutralino, and within
  simplified model scenarios. Upper limits at the 95\% confidence level are
  obtained for cross sections as functions of the masses of the
  intermediate supersymmetric particles.
}

\hypersetup{
pdfauthor={CMS Collaboration},
pdftitle={Search for supersymmetry with photons in pp collisions at sqrt(s) = 8 TeV},
pdfsubject={CMS},
pdfkeywords={CMS, physics, supersymmetry, SUSY, search, photon}}

\maketitle

\section{Introduction}

Supersymmetry (SUSY)~\cite{Ramond,Golfand,Ferrara:1974pu,Wess,Chamseddine,Barbieri,Hall}
is a popular extension of the standard model, which offers a solution to the
hierarchy problem~\cite{Barbieri198863} by introducing a supersymmetric partner
for each standard model particle.
In models with conserved $R$-parity~\cite{Barbier:2004ez,rparity}, as are
considered here, SUSY particles are produced in pairs and the lightest
supersymmetric particle (LSP) is stable.
If the LSP is weakly interacting, it escapes without detection, resulting in
events with an imbalance \ptvecmiss in transverse momentum.
Models of SUSY with gauge-mediated symmetry
breaking~\cite{GGMa,GGMd2,GGMd3,GGMd4,GGMd5,GGMd1,GGMd} predict that the
gravitino (\sGra) is the LSP.
If the next-to-lightest SUSY particle is a neutralino ($\PSGczDo$) with a bino
or wino component, photons with large transverse momenta (\pt) may be produced
in $\PSGczDo \to \gamma \sGra$ decays.
The event contains jets if the $\PSGczDo$ originates from the cascade decay of
a strongly coupled SUSY particle (a squark or a gluino).

In this paper, we present two searches for gauge-mediated SUSY particles in
proton-proton ({$\Pp\Pp$}) collisions: a search for events with at least one
isolated high-\pt photon and at least two jets, and a search for events with at
least two isolated high-\pt photons and at least one jet.
The discriminating variables are \ETm for the single-photon analysis,
and the razor variables $\MR$ and $\Rtwo$~\cite{rogan,CMS-SUS-10-009}
for the double-photon analysis, where \ETm is the magnitude of \ptvecmiss. These studies are based on a sample of
{$\Pp\Pp$}~collision events collected with the {CMS} experiment at the {CERN} {LHC} at
a center-of-mass energy of 8\TeV. The integrated luminosity of the data sample is \lumiData.

Searches for new physics with similar signatures were previously reported by the
{ATLAS} and {CMS} collaborations at ${\sqrt{s}=7\TeV}$, using samples of data no
larger than around 5\fbinv~\cite{Aad:2012zza,Aad:2012jva,Chatrchyan:2011wc,Chatrchyan:2012bba}.
No evidence for a signal was found, and models with production cross sections
larger than $\approx 10\fbinv$ were excluded in the context of general gauge-mediation~(GGM)
SUSY scenarios~\cite{GGMe,GGMf,Ruderman:2011vv,Kats:2011qh,Kats:2012ym,Grajek:2013ola}.

This paper is organized as follows. In Section~\ref{sec:detector} we
describe the CMS detector, in Section~\ref{sec:model} the benchmark signal models,
and in Section~\ref{sec:evtreco} the part of the event
reconstruction strategy that is common to the two analyses. The
specific aspects of the single- and double-photon searches are
discussed in Sections~\ref{sec:singlegamma} and~\ref{sec:razor},
respectively. The results of the analyses are
presented in Section~\ref{sec:results}. A summary is given in
Section~\ref{sec:summary}.

\section{CMS detector}
\label{sec:detector}

The central feature of the CMS apparatus is a superconducting solenoid of
6\unit{m} internal diameter, providing a magnetic field of 3.8\unit{T}.
Within the superconducting solenoid volume are a silicon pixel and strip
tracker, a lead tungstate crystal electromagnetic calorimeter (ECAL),
and a brass and scintillator hadron calorimeter (HCAL), each composed of a
barrel and two endcap sections.
Muons are measured in gas-ionization detectors embedded in the steel
flux-return yoke outside the solenoid.
Extensive forward calorimetry complements the coverage provided by the barrel
and endcap detectors.

Events are recorded using a trigger that requires the presence of at least
one high-energy photon.
This trigger is utilized both for the selection of signal events, and for the
selection of control samples used for the background determination.
The specific trigger requirements for the two analyses
are described below.
Corrections are applied to account for trigger inefficiencies,
which are evaluated using samples of data collected with orthogonal trigger conditions.
A more detailed description of the CMS detector, together with a definition of
the coordinate system used and the relevant kinematic variables, can be found
in Ref.~\cite{Chatrchyan:2008zzk}.

\section{SUSY benchmark models}
\label{sec:model}

\begin{figure*}[tbh]
  \centering
  \includegraphics[width=.495\textwidth]{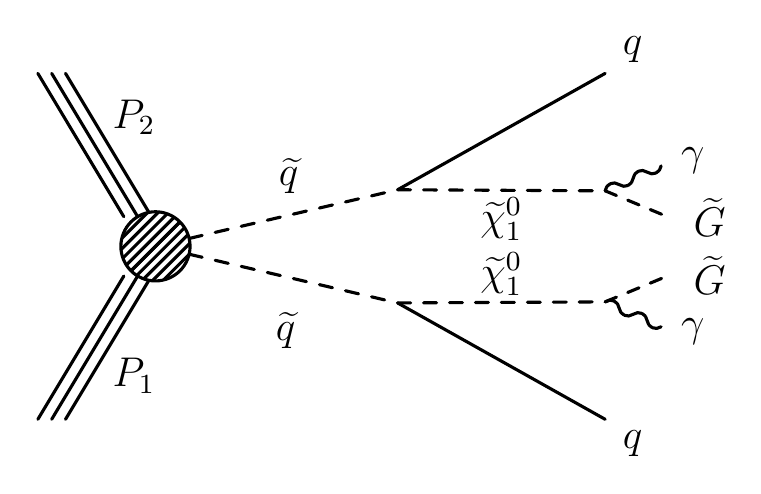}
  \includegraphics[width=.495\textwidth]{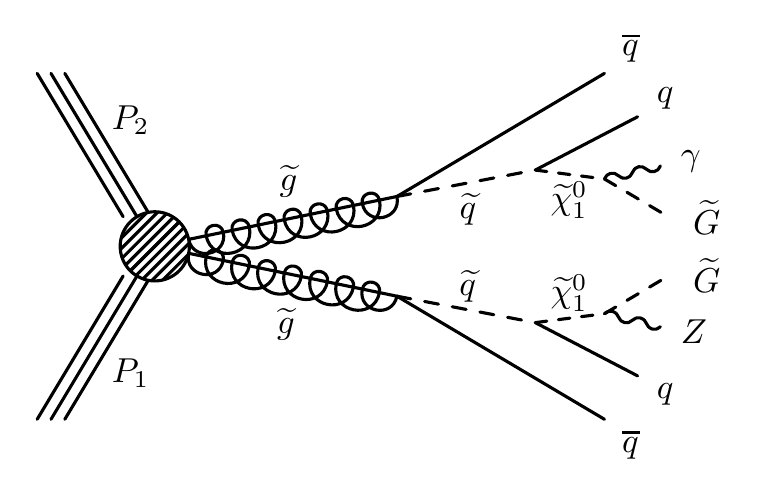} \\
  \includegraphics[width=.495\textwidth]{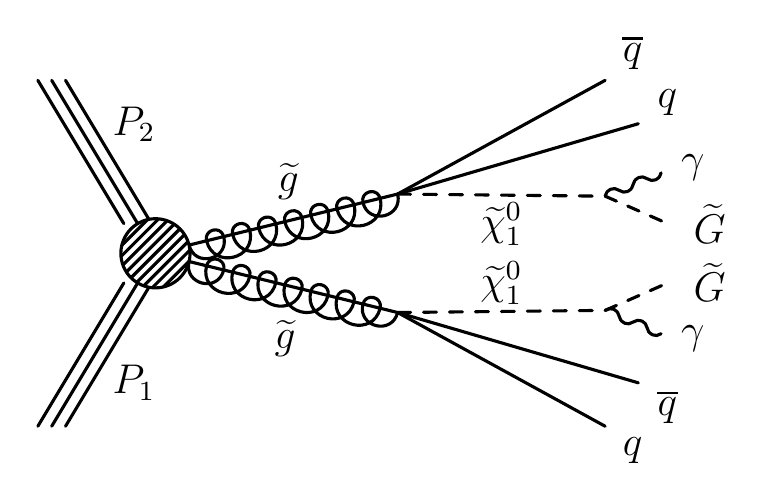}
  \includegraphics[width=.495\textwidth]{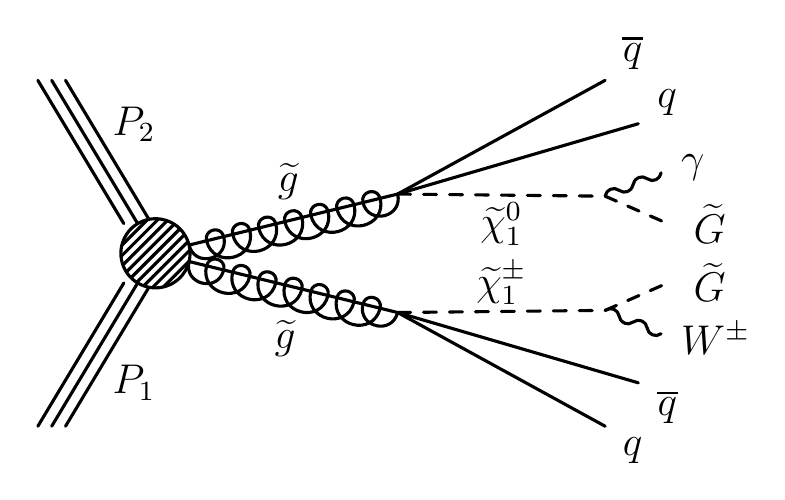}
  \caption{Typical Feynman diagrams for the general gauge-mediation model with
    bino- (top left) and wino-like (top right) neutralino mixing scenarios.
    Here, the $\PSGczDo$ can decay to $\PXXSG\gamma$ or $\PXXSG\Z$, with the
    branching fraction dependent on the $\PSGczDo$ mass.
    The diagrams for the T5gg (bottom left) and T5wg (bottom right) simplified model spectra are also shown.
    \label{fig:feynDia}
  }
\end{figure*}

The two searches are interpreted in the context of GGM SUSY
scenarios~\cite{GGMe,GGMf,Ruderman:2011vv,Kats:2011qh,Kats:2012ym,Grajek:2013ola},
and in terms of simplified model spectra (SMS) scenarios
~\cite{Alwall:2008ag,Alwall:2008va,Alves:2011sq, Alves:2011wf}
inspired by GGM models.
In these scenarios, $R$-parity is conserved and the LSP is a gravitino
with negligible mass. Four models are considered:
\begin{description}
\item[GGMbino model] In this model, squarks (\PSQ) and gluinos (\PSg) are produced and decay
    to a final state with jets and a bino-like $\PSGczDo$.
    This production process dominates over electroweak production in the
    squark- and gluino-mass region accessible to the analyses.
    The $\PSGczDo$ mass is set to 375\GeV, leading to a $\PSGczDo \to \PXXSG \gamma$
    branching fraction of about 80\%~\cite{Ruderman:2011vv}.
    The events are examined as a function of the squark and gluino masses.
    All other SUSY particles have masses set to 5\TeV,
    which renders them too heavy to participate in the interactions.
    In most cases, the final state contains two photons, jets, and \ETm.

\item[GGMwino model] This model is similar to the GGMbino model,
    except that it contains mass-degenerate wino-like $\PSGczDo$ and $\PSGcpmDo$ particles
    instead of a bino-like $\PSGczDo$.
    The common mass of the $\PSGczDo$ and $\PSGcpmDo$ is set to 375\GeV.
    The final state contains a $\gamma\gamma$, $\gamma\cPV$, or $\cPV\cPV$
    combination in addition to jets and \MET, where \cPV is a \Z or \PW~boson.
    With a $\PSGczDo \to \PXXSG \gamma$ branching fraction of about 28\%,
    approximately 48\% of all events contain at least one photon.

\item[T5gg model] This SMS model is based on gluino pair production, with the gluinos
    undergoing a three-body decay to $\cPq \bar{\cPq} \PSGczDo$, followed by $\PSGczDo \to \PXXSG
    \gamma$. All decays occur with a branching fraction of 100\%.
    The final state contains at least two photons, jets, and \ETm.

\item[T5wg model] This SMS model is also based on gluino pair production, with one gluino
    undergoing a three-body decay to $\cPq \cPaq \PSGczDo$, followed by $\PSGczDo \to \PXXSG
    \gamma$, and the other gluino undergoing a three-body decay to $\cPq \cPaq \PSGcpmDo$,
    followed by $\PSGcpmDo \to \PXXSG \Wpm$.
    All decays occur with a branching fraction of 100\%.
    The final state contains at least one photon, jets, and \ETm.
\end{description}
Typical Feynman diagrams corresponding to these processes are shown in Fig.~\ref{fig:feynDia}.
Note that for the two GGM models, the events can proceed through the production
of gluino-gluino, gluino-squark, or squark-squark pairs.

Signal events for the GGM models are simulated
using the \PYTHIA6~\cite{Sjostrand:2006za} event generator.
The squark and gluino masses are varied between 400 and 2000\GeV. Eight
mass-degenerate squarks of different flavor (\cPqu, \cPqd, \cPqs, and \cPqc) and chirality
(left and right) are considered. The production cross sections are normalized to
next-to-leading order (NLO) in quantum chromodynamics, determined using the \PROSPINO~\cite{NLONLL1} program, and is
dominated by gluino-gluino, gluino-squark, and squark-squark production.

The SMS signal events are simulated with the \MADGRAPH5~\cite{Alwall:2011uj} Monte Carlo (MC) event generator in association
with up to two additional partons.
The decays of SUSY particles, the parton showers, and the hadronization of partons,
are described using the \PYTHIA6 program.
Matching of the parton shower with the \MADGRAPH5 matrix element calculation is
performed using the MLM~\cite{Alwall:2007fs} procedure. The gluino pair-production
cross section is described to NLO+NLL
accuracy~\cite{NLONLL1,NLONLL2,NLONLL3,NLONLL4,NLONLL5}, where NLL refers to
next-to-leading-logarithm calculations.  All SUSY particles except
the gluino, squark, LSP, and $\PSGc_1$ states are assumed to be too
heavy to participate in the interactions. The NLO+NLL cross section and
the associated theoretical uncertainty~\cite{NLONLLerr} are taken as a
reference to derive exclusion limits on SUSY particle masses. Gluino masses
of 400 (800) to 1600\GeV, and $\PSGc_1$ masses up to 1575\GeV, are probed
in the T5wg (T5gg) model.

For all the signal models, detector effects are simulated through a
fast simulation of the CMS experiment~\cite{CMS-DP-2010-039}.

\section{Event reconstruction}
\label{sec:evtreco}

The events selected in this study are required to have at least one
high quality reconstructed interaction vertex.
The primary vertex is defined as the one with the highest sum of the $\pt^2$ values of the associated tracks.
A set of detector-
and beam-related noise cleaning algorithms is applied to remove events with
spurious signals, which can mimic signal events with high energetic particles
or large \ETm~\cite{CMS-JME-10-009,CMS-SUS-12-011}.

Events are reconstructed using the particle-flow
algorithm~\cite{CMS-PAS-PFT-09-001,CMS-PAS-PFT-10-001},
which combines information from various detector components
to identify all particles in the event.
Individual particles are
reconstructed and classified in five categories: muons, electrons,
photons, charged hadrons, and neutral hadrons. All neutral particles, and
charged particles with a track pointing to the primary vertex, are
clustered into jets using the anti-\kt clustering
algorithm~\cite{Cacciari:2008gp}, as implemented in the \FASTJET package~\cite{fastjet},
with distance parameter of 0.5. The momenta of the jets are corrected for the response
of the detector and for the effects of multiple interactions in the
same bunch crossing (pileup)~\cite{Cacciari:2007fd}.
Jets are required to satisfy loose quality criteria that remove candidates caused by detector noise.

Photons are reconstructed from clusters of energy in the ECAL~\cite{Khachatryan:2015iwa}.
The lateral distribution of the cluster energy is required to be consistent with that
expected from a photon, and the energy detected in the HCAL behind the
photon shower cannot exceed 5\% of the ECAL cluster energy.  A veto is
applied to photon candidates that match hit patterns consistent with a track
in the pixel detector (pixel seeds), to reduce spurious photon candidates originating from electrons.
Spurious photon candidates originating from quark and gluon
jets are suppressed by requiring each photon candidate to be isolated
from other reconstructed particles. In a cone of radius
$\Delta R \equiv \sqrt{ \smash[b]{(\Delta\eta)^2+(\Delta\phi)^2 }} = 0.3$ around the
candidate's direction, the scalar \pt sums of charged hadrons~(\ciso), neutral
hadrons~(\niso), and other electromagnetic objects~(\piso) are
separately formed, excluding the contribution from the candidate
itself.  Each momentum sum is corrected for the pileup contribution,
computed for each event from the estimated energy density in the
$(\eta,\phi)$ plane.
Selected photons are required to be isolated according to criteria imposed
on \ciso, \niso, and \piso as defined in Ref.~\cite{Khachatryan:2015iwa}.

\section{Single-photon search}
\label{sec:singlegamma}

The single-photon analysis is based on a trigger requiring the presence of at least one
photon candidate with $\pt\geq70\GeV$.
The trigger also requires $\HT>400\GeV$, where $\HT$ is the scalar sum of jet
$\pt$ values for jets with $\pt\geq40\GeV$ and $\abs{\eta}\leq3$,
including photons that are misreconstructed as jets.

In the subsequent analysis, we make use of the variable \ptStar, which is
defined by considering the photon candidate and nearby reconstructed particles,
clustered as a jet as described in Section~\ref{sec:evtreco}. If a jet
(possibly including the photon) is reconstructed within $\Delta R<0.2$ of the
photon candidate and the \pt value of the jet is less than three times of that
of the photon candidate itself, it is referred to as the ``photon jet''. If
such a jet is found, \ptStar is defined as the \pt value of the photon jet.
Otherwise, \ptStar is the \pt value of the photon candidate.
We require photon candidates to satisfy $\ptStar>110\gev$ and $\abs{\eta}<1.44$.
Also, in the subsequent analysis, we make use of the
variable \HTstar, defined as for \HT in the previous paragraph but
including the \ptStar values of all selected photon candidates.
The variables \ptStar and \HTstar reduce differences between photon candidates
selected with different isolation requirements compared to the unmodified variables \pt and \HT.
We require events to satisfy $\HTstar\geq 500\GeV$.
The sample of events with isolated photons so selected is referred to as the \ggamma sample.
The trigger efficiency for the selected events to enter the sample is
determined to be 97\%, independent of $\ptStar$ and $\HTstar$.

We require at least two jets with $\pt\geq30\GeV$ and $\abs{\eta}\leq2.5$.
The jets must be separated by $\Delta R \geq 0.3$ from all photon candidates, to prevent double counting.
In addition, the requirement $\ETm\geq100\GeV$ is imposed and events with isolated
electrons or isolated muons are vetoed.
The selection is summarized in
Table~\ref{tab:cutsummary}.
Note that 0.16\% of the selected events contain more than one photon candidate.
\begin{table}[htbp]
  \topcaption{Summary of the single-photon analysis selection criteria.
          \label{tab:cutsummary}}

  \centering
  \begin{scotch}{l|c|c|c}
    \multirow{2}{*}{ Selection criteria } & $\ggamma$     & $\fgamma$      & $\egamma$ \\
                                          & signal reg. & control reg. & control reg. \\
    \hline
    Isolation requirement                 & Tight         &  Loose         & Tight \\
    Pixel seed                            & Vetoed        & Vetoed         & Required \\
    \hline
    \multirow{3}{*}{Trigger} & \multicolumn{3}{c}{ $\gamma$-\HT trigger with }\\
    & \multicolumn{3}{c}{ $\pt^\gamma\geq70\GeV$, $\HT\geq400\GeV$ }\\
    & \multicolumn{3}{c}{ (using $\pt^\text{jets}\geq40\GeV$, $\abs{\eta}\leq3.0$) }\\
    \hline
    Photon(s) & \multicolumn{3}{c}{ $\geq$1, ${\ptStar}^{\gamma}\geq110\GeV$, $\abs{\eta}\leq1.44$ }\\
    \hline
    Jet(s) & \multicolumn{3}{c}{ $\geq$2, $\pt^{\text{jets 1,2}}\geq30\GeV$, $\abs{\eta}\leq2.5$ }\\
    \hline
    \multirow{2}{*}{\HTstar} & \multicolumn{3}{c}{ $\geq 500\GeV$ }\\
    & \multicolumn{3}{c}{ (using $\pt^{\text{jets, $\gamma$}}\geq40\GeV$, $\abs{\eta}\leq3.0$) }\\
    \hline
    Isolated \Pe,$\mu$ & \multicolumn{3}{c}{ veto, $\pt \geq 15\GeV$, $\abs{\eta^{\Pe(\mu)}}\leq 2.5\,(2.4)$}\\
    \hline
    \ETm & \multicolumn{3}{c}{ $\ETm\geq 100\GeV$ (six ranges in \ETm) }\\
    \end{scotch}
\end{table}

The relevant sources of background to the single-photon search are:
\begin{itemize}
  \item multijet events
    with large \ETm, originating from the mismeasured momenta of some of
    the reconstructed jets. This class of events contains both genuine photons
    and spurious photon candidates from jets. This is by far the largest
    contribution to the background.
  \item events with
    genuine \ETm originating from the leptonic decay of \PW~bosons, both
    directly produced and originating from top quark decays,
    which we refer to as electroweak (EW) background.
  \item rare processes with initial- or final-state photon
    radiation (ISR/FSR), such as $\gamma\PW$, $\gamma\Z$ (especially $\gamma\Z\to\gamma\nu\nu$),
    and $\gamma\ttbar$ production.
\end{itemize}

\begin{figure}[ht!bp]
  \centering
  \includegraphics[width=0.48\textwidth]{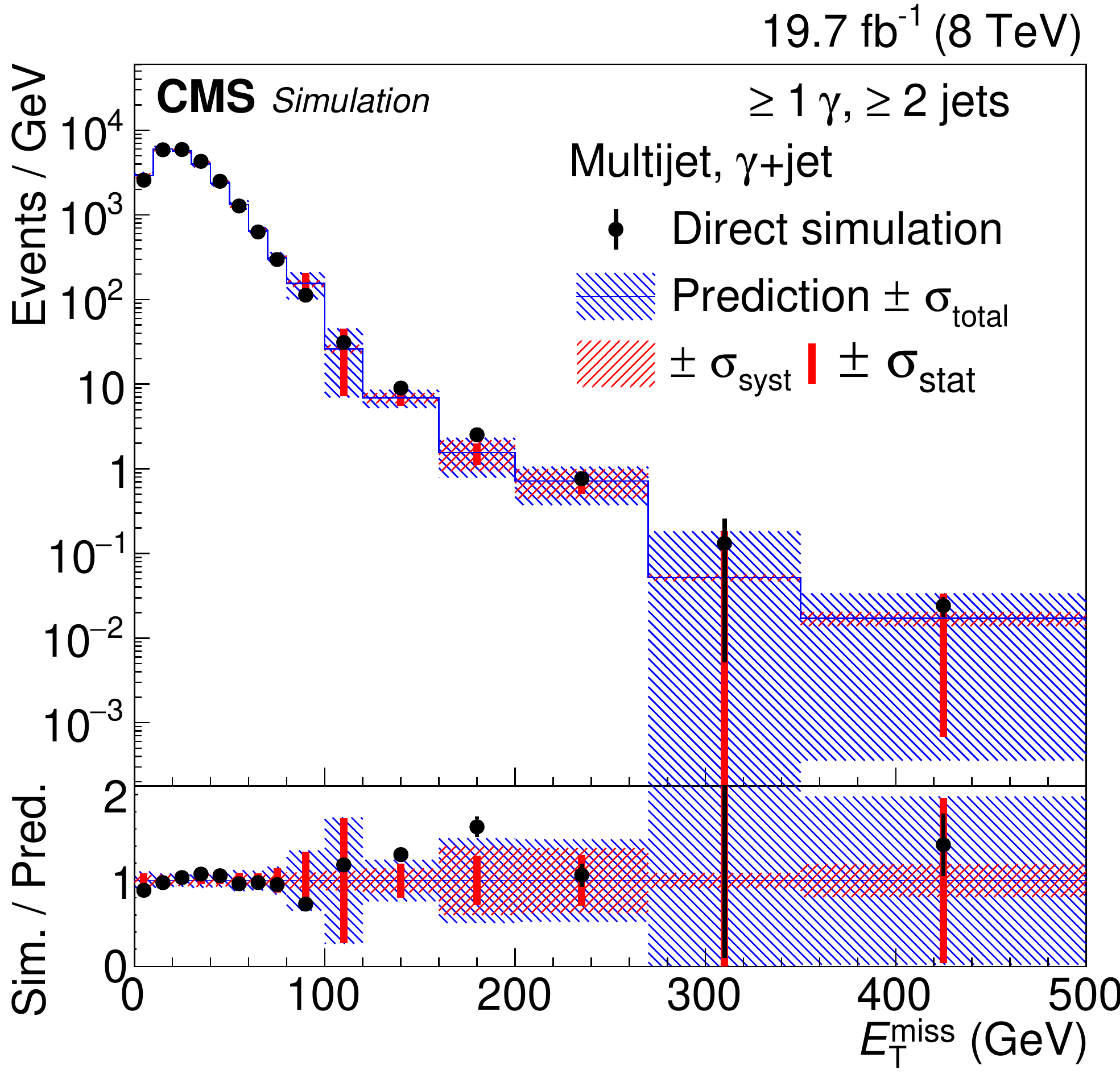}
  \includegraphics[width=0.48\textwidth]{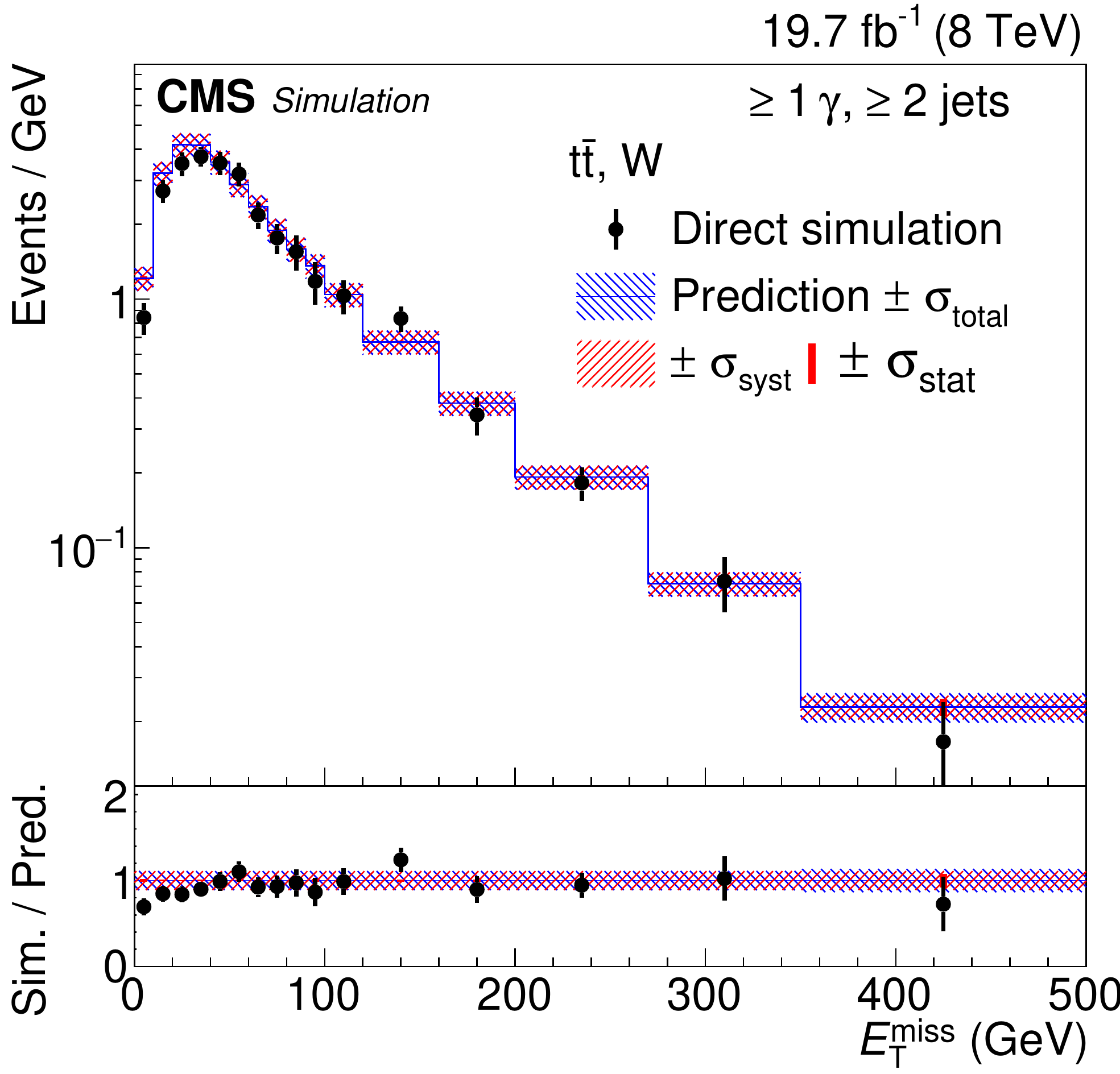}
  \caption{
    Tests of the background estimation method for the single-photon analysis
    using simulated events in bins of \ETm. The direct simulation of
    \ggamma events is compared to the prediction of the multijet background from
    simulated \fgamma events (\cmsLeft).
    Simulated events with \ggamma originating from generated electrons are compared
    to the simulated prediction using the EW background method (\cmsRight).
    The blue hatched area represents the total
    uncertainty and is the quadratic sum of the statistical (red vertical bars)
    and systematic (red hatched area) uncertainties.  In the bottom panels, the
    ratio of the direct simulation to the prediction is shown.
    \label{fig:ClosureTests}
  }
\end{figure}

The kinematic properties of the multijet background are estimated from a
control sample of photon candidates with isolation-variable values
(\ciso, \niso, \piso) too large to satisfy the signal photon selection.
We refer to these events as the $\fgamma$ sample. Photon candidates of this
kind typically originate from jets with anomalous fractions of energy
deposited in the ECAL. Other than the orthogonal requirement of a \fgamma rather
than a \ggamma candidate, events in this control sample are selected
with the same requirements as the \ggamma sample, as summarized in
Table~\ref{tab:cutsummary}. Despite the different isolation
requirement, this sample has properties similar to those of the \ggamma
sample, due to the use of \ptStar rather than photon \pt in the
definition of the event kinematic variables. Moreover, events in the
\fgamma control sample are corrected for a residual difference with respect to the \ggamma sample in the
distributions of \ptStar and hadronic recoil \pt, estimated from
events with $\ETm<100\GeV$. The corrected distribution of a given
kinematic property (\eg, \ETm) for \fgamma events provides an
estimate of the corresponding distribution for \ggamma events. The
uncertainty in the correction factors, propagated to the prediction, is fully
correlated among bins in the signal region and is treated as
a systematic uncertainty in the background yield. The limited statistical precision of the control sample dominates the total systematic
uncertainty.
Figure~\ref{fig:ClosureTests} (\cmsLeft) shows the \ETm distribution
from the \ggamma sample and the corresponding prediction from the \fgamma sample,
for simulated multijet and $\gamma+\text{jet}$ events.
No discrepancy is observed within the
quoted uncertainties.

The EW background is characterized by the presence of an electron
misidentified as a photon. The kinematic properties of this background
are evaluated from a second control sample, denoted the \egamma sample, defined by
requiring at least one pixel seed matching the photon candidate but otherwise
using the \ggamma selection criteria, as summarized in
Table~\ref{tab:cutsummary}.
Events in the \egamma sample are weighted by the probability \fakeRate for an
electron to be misidentified as a photon, which is measured as a function of
the $\gamma$ candidate \pt, the number of tracks associated with the primary vertex,
and the number of reconstructed vertices in an event by determining the rate of
events with reconstructed $\Pe\egamma$ and $\Pe\ggamma$ combinations in a sample of
$\Z\to\EE$ events.
The event-by-event
misidentification rate is about 1.5\%, with a weak
dependence on the number of vertices.
A systematic uncertainty of 11\% is assigned to \fakeRate to account for the
uncertainty in the shape of the function and for differences between the control sample in which the misidentification
rate is calculated and the control sample to which it is applied.
The predicted \ETm distribution for the EW background,
obtained from a simulated sample of \PW~boson and \ttbar events, is
shown in Fig.~\ref{fig:ClosureTests} (\cmsRight) in comparison with the results
from the direct simulation of events with \ggamma originating
from electrons.  The distributions agree within the quoted
uncertainties.

The contribution of ISR/FSR background events is estimated from simulation
using leading-order results from the \MADGRAPH5 MC
event generator with up to two additional partons, scaled by a factor of $1.50\pm0.75$
including NLO corrections determined with the \MCFM~\cite{Campbell:2012ft,CMS-EXO-12-047}
program.

\begin{figure}[tb]
  \centering
  \includegraphics[width=0.48\textwidth]{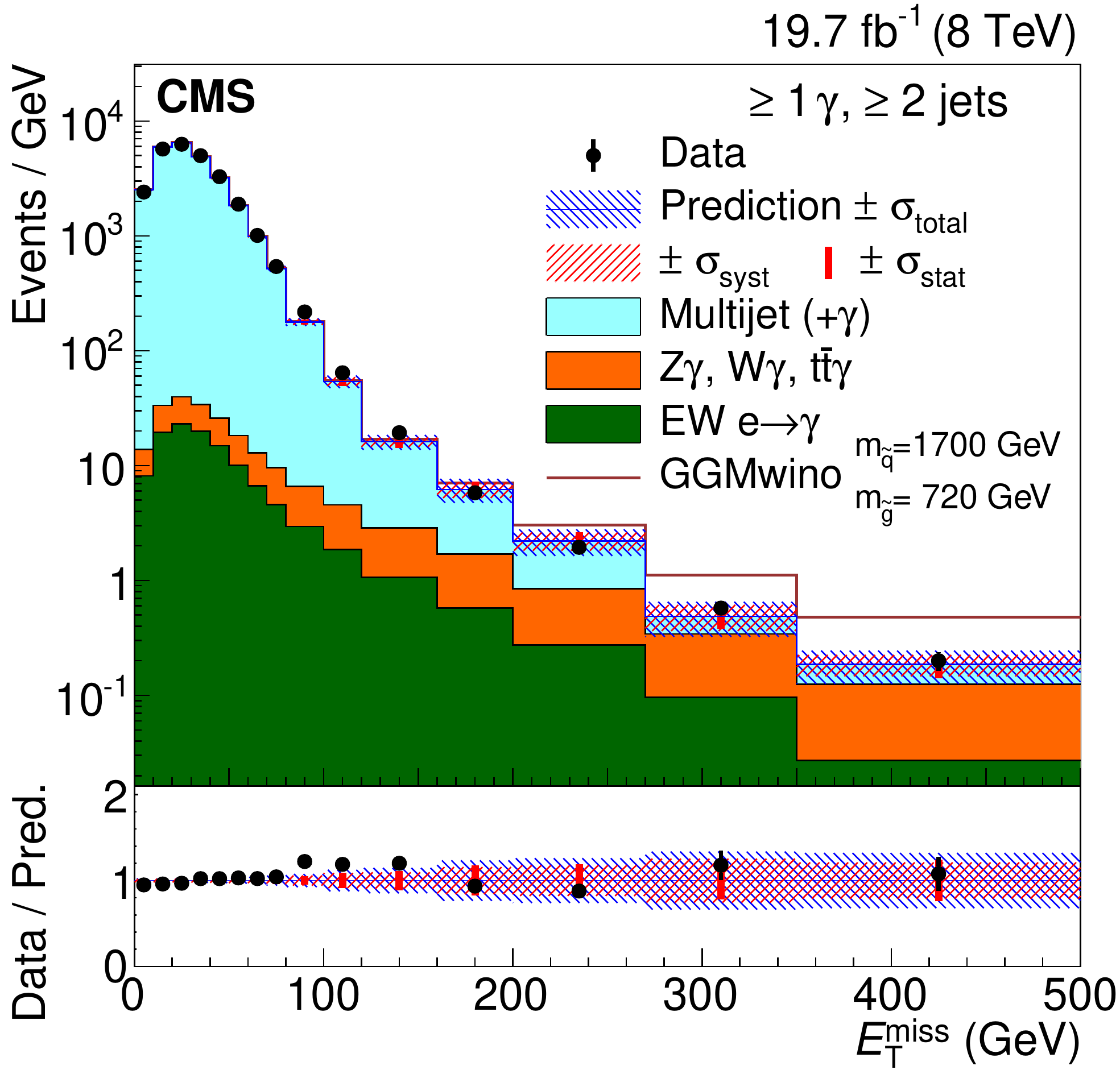}
  \caption{Distribution of \ETm from the single-photon search in comparison
    to the standard model background prediction. The expectation
    from an example GGMwino signal model point is also shown.
    In the bottom panels, the ratio of the data to the prediction is shown.
    The representations of uncertainties are defined as in Fig.~\ref{fig:ClosureTests}.
    \label{fig:preselmet}
  }
\end{figure}

The measured \ETm spectrum in the \ggamma sample is shown in Fig.~\ref{fig:preselmet}
in comparison with the predicted standard model background.
A SUSY signal would appear as an excess at large \ETm above the
standard model expectation. Figure~\ref{fig:preselmet} includes, as an example,
the simulated distribution for a benchmark
GGMwino model with a squark mass of 1700\GeV,
a gluino mass of 720\GeV, and a total NLO cross section of 0.32\unit{pb}.

For purposes of interpretation, we divide the data into six bins of \ETm,
indicated in Table~\ref{resulttable}.  For each bin, Table~\ref{resulttable}
lists the number of observed events, the number of predicted standard model
events, the acceptance for the benchmark signal model, and the number of
background events introduced by the predicted signal contributions to the
control regions, where this latter quantity is normalized to the corresponding
signal yield.

No significant excess of events is observed.
An exclusion limit on the signal yield is derived at 95\% confidence level
(\CL), using the \CLs method~\cite{Junk:1999kv,Read,LHCCLs}.
For a given signal hypothesis, the six \ETm signal regions are combined in a
multichannel counting experiment to derive an upper limit on the
production cross section.
The results, presented in Section~\ref{sec:results}, account
for the possible contribution of signal events to the two control
samples, which lowers the effective acceptance by 10--20\% depending
on the assumed SUSY mass values.

\begin{table*}[htb]
\newcommand{\tableMSquark}{1700}
\newcommand{\tableMGluino}{720}
\newcommand{\tableSignalXSec}{0.32}
\newcommand{\tableObsLimit}{0.11}
\newcommand{\tableExpLimit}{0.13}
\topcaption{
  Observed numbers of events and standard model background
  predictions for the single-photon search.
  The signal yield and acceptance for the GGMwino model with
  $m_{\PSQ}=\tableMSquark\GeV$ and $m_{\PSg}=\tableMGluino\GeV$, with a total
  signal cross section of $\sigma_{\mbox{\footnotesize NLO}}=\tableSignalXSec\unit{pb}$,
  are also shown.
  The last line gives the additional number of background events, normalized to
  the signal yield, which is associated with signal contributions to the two
  control regions.
  \label{resulttable}
}
\centering
\begin{scotch}{lcccccc}
$\ETm$ range $(\GeVns{})$    &  [100,120) & [120,160) & [160,200) & [200,270) & [270,350) & [350,$\infty$) \\
\hline
Multijet        & $991\pm164$  & $529\pm114$   & $180\pm69\phantom{0}$      & $96\pm45$            & $12\pm12$           & $9\pm9$  \\
ISR/FSR         & $54\pm27$    & $73\pm36$ & $45\pm23$           & $40\pm20$            & $20\pm10$           & $15\pm7\phantom{0}$ \\
EW              & $37\pm4\phantom{0}$ & $43\pm5\phantom{0}$  & $23\pm3\phantom{0}$  & $19\pm2\phantom{0}$  & $8\pm1$  & $4\pm1$\\
Background      & $1082\pm166\phantom{0}$  & $644\pm119$      & $248\pm73\phantom{0}$ & $155\pm50\phantom{0}$ & $39\pm16$ & $28\pm12$\\ \hline
Data            & $1286$ & $774$ & $232$ & $136$ & $46$ & $30$         \\
Signal yield    & $19\pm3\phantom{0}$     & $53\pm5\phantom{0}$ & $51\pm5\phantom{0}$  & $82\pm7\phantom{0}$  & $78\pm7\phantom{0}$ & $67\pm6\phantom{0}$ \\
Signal acceptance [\%]     & 0.3          & 0.9         & 0.8         & 1.3         & 1.2         & 1.1        \\
\multirow{3}{20ex}{Background from signal relative to the signal yield [\%]} &&&&&&\\

& 2.1          & 5.0         & 5.6         & 9.9         & 26.7        & 13.5       \\
&&&&&&\\
\end{scotch}
\end{table*}

\section{Double-photon search}
\label{sec:razor}

Events considered for the double-photon search are collected using triggers
developed for the discovery of the Higgs boson in diphoton
events~\cite{CMS-HIG-12-028,CMS-HIG-12-036,CMS-HIG-13-001}. These triggers
use complementary kinematic selections:
\begin{itemize}
  \item two photons with $\pt>18\GeV$, where the highest \pt photon is required
    to have $\pt>26\GeV$, while the diphoton invariant mass is required to be larger than $70\GeV$.
  \item two photons with $\pt>22\GeV$, where the highest \pt photon is required to have $\pt>36\GeV$.
\end{itemize}
In addition, each photon must satisfy at least one of two requirements:
a high value of the shower shape variable $R_9$~\cite{Khachatryan:2015iwa} or loose calorimetric identification.
For the targeted signals, the combination of the two triggers is found to be 99\% efficient.

In the subsequent analysis, at least two photon candidates with $\pt > 22 \GeV$ and
$\abs{\eta}<2.5$ are required. Events are selected if the
highest \pt photon has $\pt > 30 \GeV$.
Jets must have $\pt>40 \GeV$ and $\abs{\eta}<2.5$, with each jet
required to lie a distance $\Delta R > 0.5$ from an identified photon.
Only events with at least one selected jet are
considered.

The background is dominated by multijet events, which mostly consist of events
with at least one genuine photon. Due to the requirement of two
photons in the event, the EW and ISR/FSR backgrounds are negligible.

The razor variables $\MR$ and $\Rtwo$~\cite{rogan,CMS-SUS-10-009}
are used to distinguish a potential signal from
background. To evaluate these variables, the selected jets and photons are
grouped into two exclusive groups, referred to as
``megajets''~\cite{CMS-SUS-10-009}.  The four-momentum of a megajet is
computed as the vector sum of the four-momenta of its
constituents. Among all possible megajet pairs in an event, we
select the pair with the smallest sum of squared invariant masses
of the megajets.
Although not explicitly required, the two photons are associated with
different megajets in more than 80\% of the selected signal events.

The variable $\MR$ is defined as
\begin{linenomath}
\begin{equation}
  \MR \equiv \sqrt{
    \left(\abs{\vec{p}^{\,\mathrm{j}_{1}}_{\phantom{z}}}+\abs{\vec{p}^{\,{ \mathrm{j}_{2}}}_{\phantom{z}}}\right)^2 -
    \left({p}^{\mathrm{j}_1}_z+{p}^{\mathrm{j}_2}_z\right)^2
  },
\end{equation}
\end{linenomath}
where $\vec{p}^{\,\mathrm{j}_i}_{\phantom{z}}$ and
$p^{\mathrm{j}_i}_z$ are, respectively, the momentum of
the $i$th megajet and the magnitude of its component along the beam axis.
The \pt imbalance in the
event is quantified by the variable $\MRT$, defined as
\begin{linenomath}
\begin{equation}
  \MRT \equiv \sqrt{ \frac{
    \ETm \left(\abs{\ptvec^{\,\mathrm{j}_1}}+\abs{\ptvec^{\,\mathrm{j}_2}}\right) -
    \ptvecmiss \cdot  \left(\ptvec^{\,\mathrm{j}_1}+\ptvec^{\,\mathrm{j}_2}\right)}{2}},
\end{equation}
\end{linenomath}
where $\ptvec^{\,\mathrm{j}_i}$ is the transverse component of $\vec{p}^{\,\mathrm{j}_i}_{\phantom{z}}$.
The razor ratio \R is defined as
\begin{linenomath}
\begin{equation}
  \R \equiv \frac{\MRT}{\MR}.
\end{equation}
\end{linenomath}

\begin{figure}[tbh]
  \centering
  \includegraphics[width=.495\textwidth]{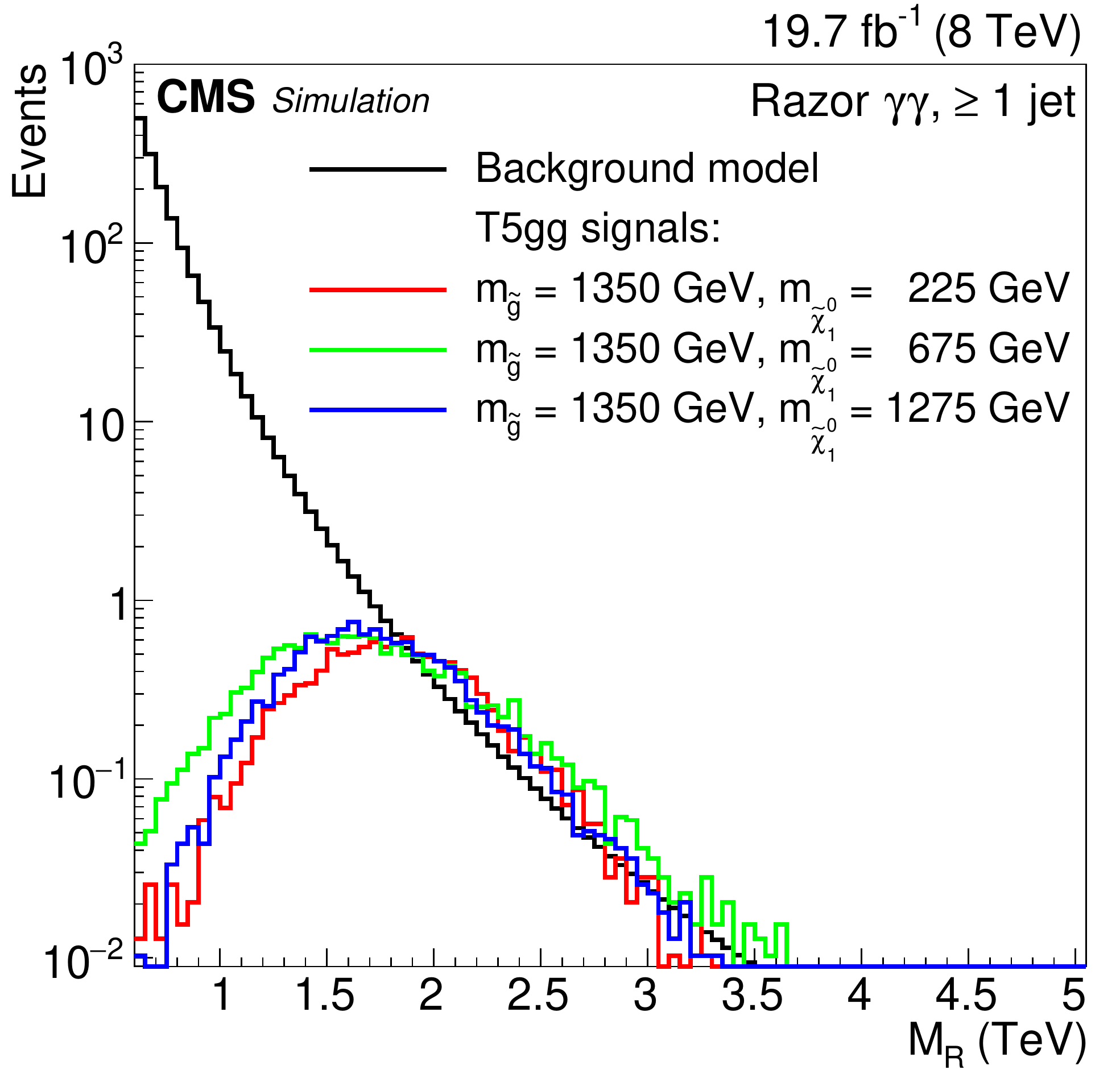}
  \includegraphics[width=.495\textwidth]{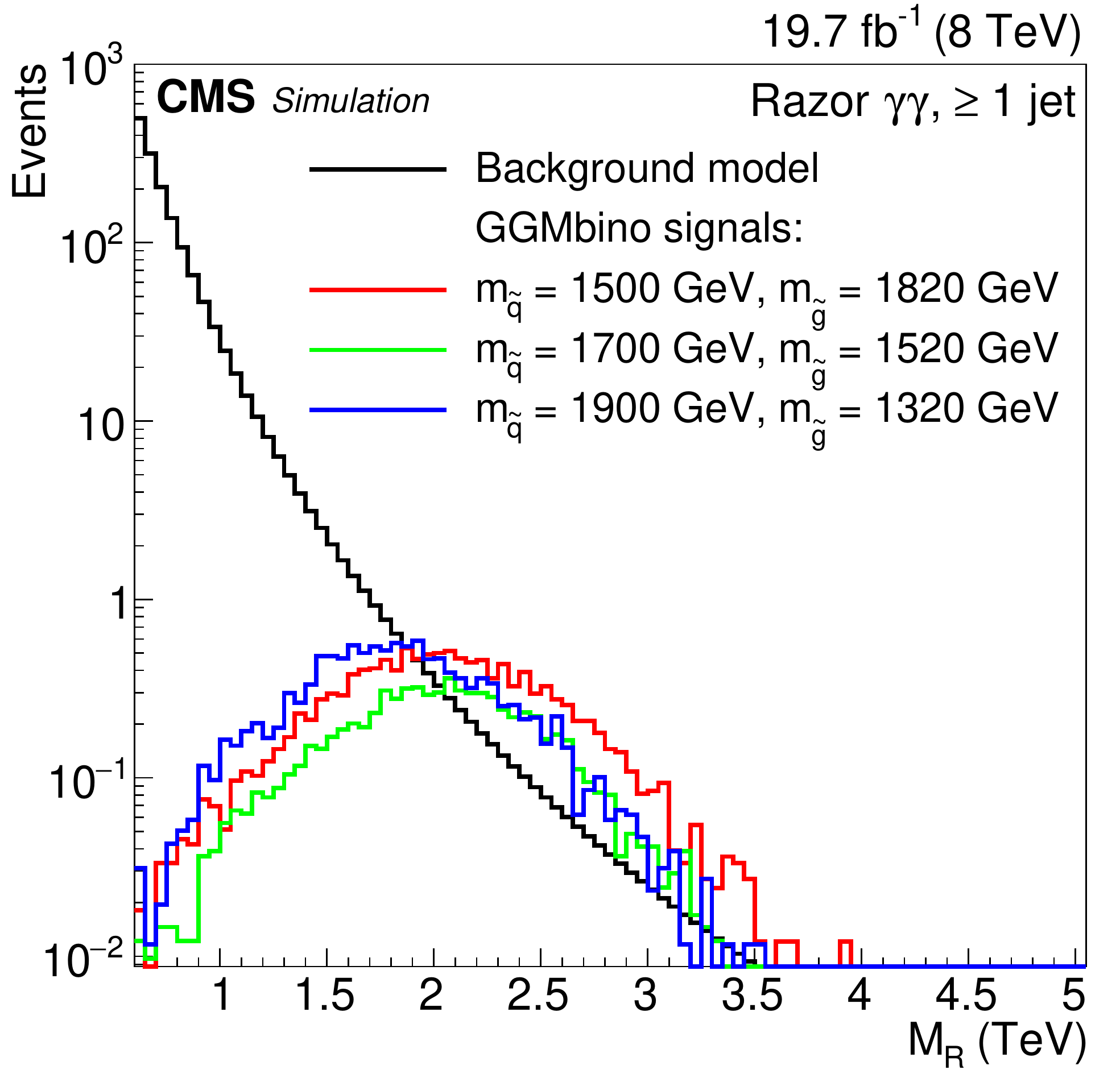}
  \caption{
    Distribution of $\MR$ in the double-photon search for the background model,
    derived from a fit in the data control region,
    and for the T5gg (\cmsLeft) and GGMbino (\cmsRight) signal models.
    The background model is
    normalized to the number of events in the signal region. The signal
    models are normalized to the expected signal yields.
    \label{fig:ffFit2}
  }
\end{figure}

For squark pair production in $R$-parity conserving models in which both
squarks decay to a quark and LSP, the $\MR$ distribution peaks at
$M_\Delta = (m_{\PSq}^2 - m_{\mathrm{LSP}}^2)/m_{\PSq}$,
where $m_{\PSQ}$ $(m_{\mathrm{LSP}})$ is the squark (LSP) mass.
Figure~\ref{fig:ffFit2} demonstrates that $\MR$ also peaks for gluino pair
production (\cmsLeft) and in the GGMbino model (\cmsRight).

The ($\MR$, $\Rtwo$) plane is divided into two regions:
(i) a signal region with $\MR> 600 \GeV$
and $\Rtwo>0.02$, and (ii) a control region with
${\MR> 600 \GeV}$ and ${0.01<\Rtwo\leq 0.02}$. The control
region is defined such that any potential signal contribution to the control region is less
than 10\% of the expected number of signal events,
producing a negligible bias on the background shape determination,
corresponding to less than a 2\% shift in the predicted number of background
events for 20 expected signal events.

The background shape is determined through a maximum likelihood fit of
the $\MR$ distribution in the data control region, using the
empirical template function
\begin{linenomath}
\begin{equation}
  P(\MR) \propto \re^{-k\left(\MR - \MRz\right)^{\frac{1}{n}}},
\end{equation}
\end{linenomath}
with fitted parameters $k$, \MRz, and $n$.
The best-fit shape is used to describe the $\MR$ background distribution
in the signal region, fixing the overall normalization to the observed
yield in the signal region.
This implicitly assumes a negligible contribution of
signal events to the overall normalization.  We have studied the impact of
the resulting bias and found it to be negligible for the expected signal
distributions and magnitudes.
The covariance matrix derived from the fit
in the control region is used to sample an ensemble of alternative
$\MR$ background shapes.  For each bin of the $\MR$
distribution, a probability distribution for the yield is derived using
pseudoexperiments. The uncertainty in each bin is defined by requiring 68\%
of the pseudoexperiments to be contained within the uncertainty band.

\begin{figure}[tbhp]
  \centering
  \includegraphics[width=.495\textwidth]{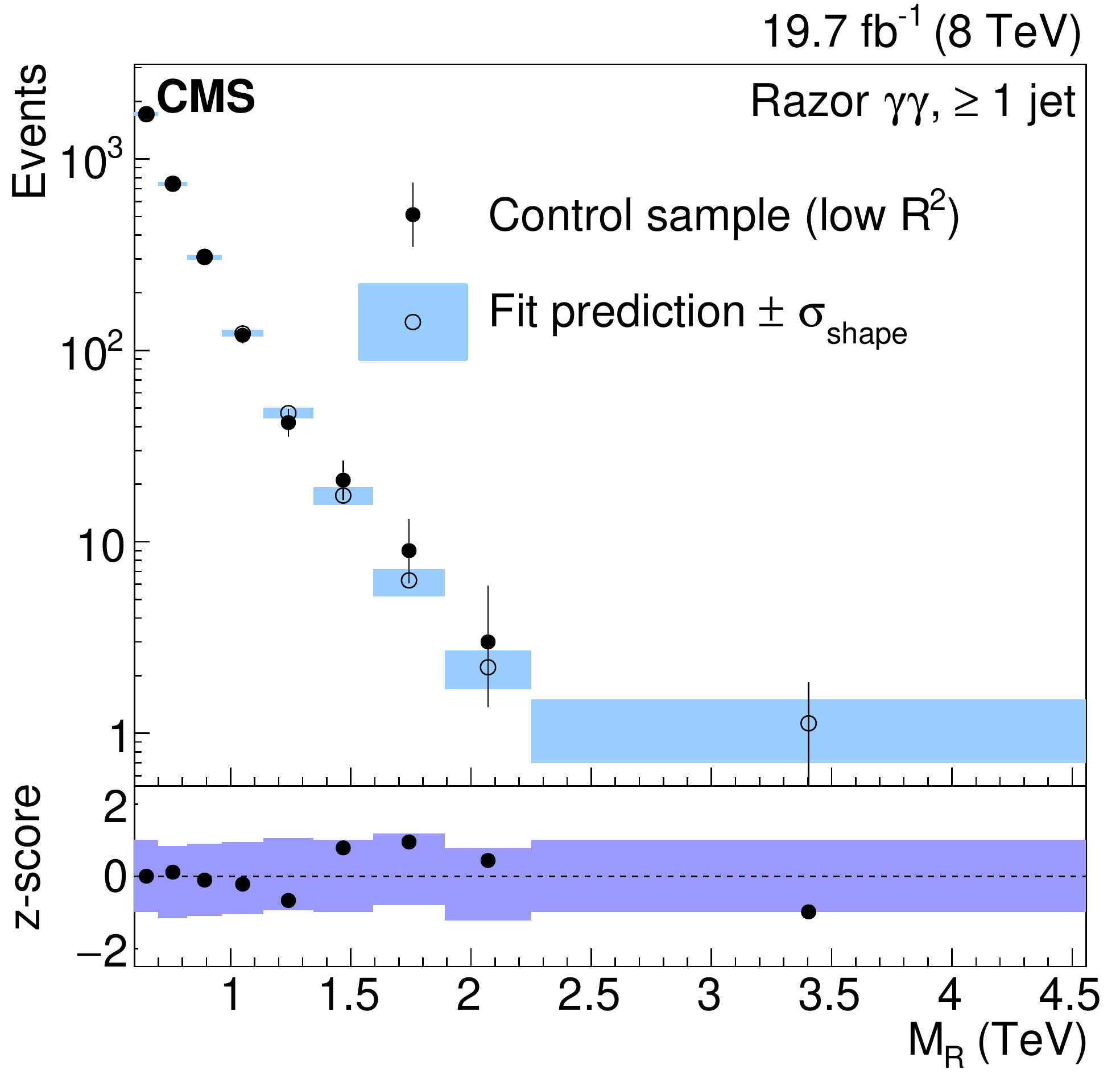}
  \includegraphics[width=.495\textwidth]{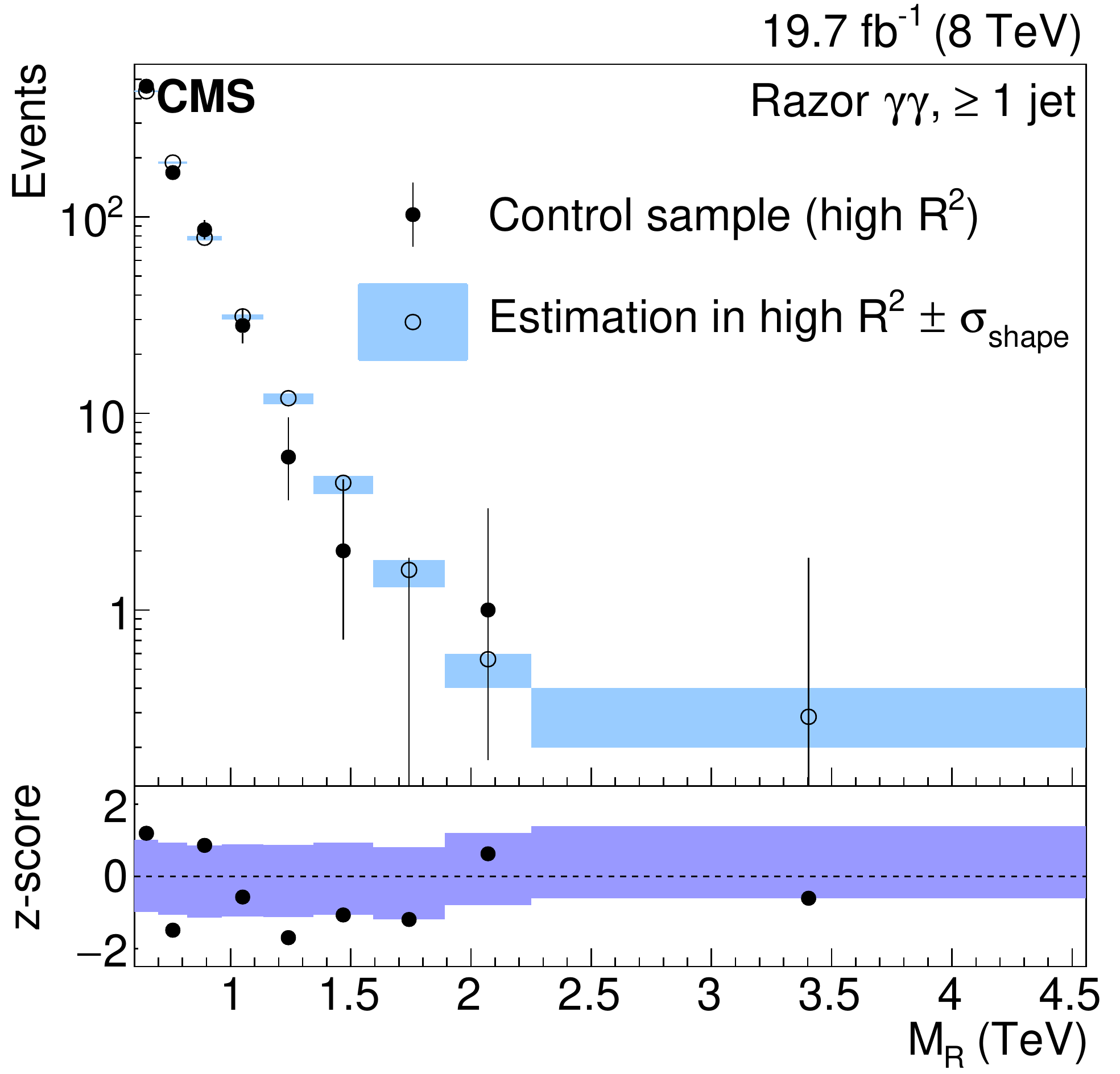}
  \caption{Distribution of $\MR$ in the double-photon search for a control sample of jets misreconstructed
    as photons (see text) in the control (\cmsLeft) and signal (\cmsRight) regions.
    The data are compared to the 68\% range obtained from a fit in the control
    region and extrapolated to the signal region (blue bands).
    The open dots represent the center of the 68\% range.
    The rightmost bin in each plot contains zero data entries.
    The bottom panel of each figure gives the z-scores (number of Gaussian
    standard deviations) comparing the filled dots to the band.
    The filled band shows the position of the 68\% window with respect to the
    expected value.
    \label{fig:ffFit}
  }
\end{figure}

This background prediction method is tested by applying it to a control sample of events in which
jets are misidentified as photons, obtained by
selecting photon candidates that fail the requirement on the cluster
shape or the photon isolation. The remainder of the photon-selection criteria are the same as for the signal sample.
In Fig.~\ref{fig:ffFit} we show the fit
result in the control region (\cmsLeft) and the extrapolation to the signal
region (\cmsRight).

The contribution of the EW and ISR/FSR backgrounds, characterized
by genuine \ETm, is evaluated from simulated events and is found to be
negligible compared to the systematic uncertainty associated with the multijet background method,
and is accordingly ignored.

\begin{figure}[tbh]
  \centering
  \includegraphics[width=.495\textwidth]{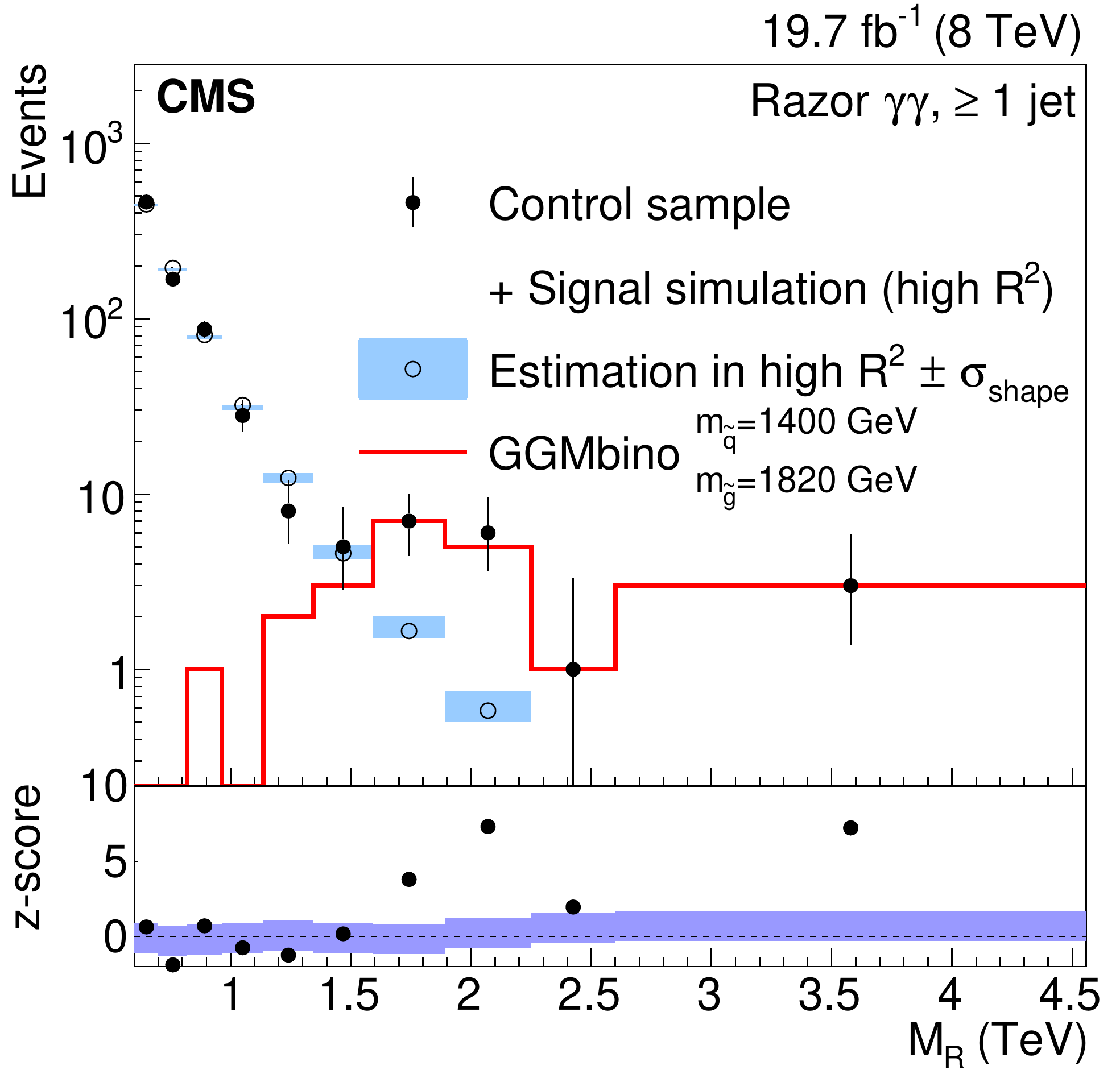}
  \caption{Distribution of $\MR$ in the double-photon search for a control sample of jets
    misreconstructed as photons to which a simulated sample of GGMbino
    events has been added. The squark and gluino masses are respectively set to $m_\PSq =
    1400\GeV$ and $m_\PSg = 1820 \GeV$, and the production cross
    section is fixed to $\sigma = 2.7\unit{fb}$. The signal contribution is
    shown by the red histogram.
    The representations of the uncertainty bands, data points, and the information shown in the bottom panel are the same as in
    Fig.~\ref{fig:ffFit}.
    \label{fig:signalInjection}
  }
\end{figure}

A signal originating from heavy squarks or gluinos would result in a
wide peak in the $\MR$ distribution. This is shown in
Fig.~\ref{fig:signalInjection}, where a GGMbino signal sample
is added to the control sample of jets misreconstructed as photons,
and the background prediction
method is applied. The contribution of signal events to the control region is
negligible and does not alter the background
shape of Fig.~\ref{fig:ffFit} (\cmsLeft).
The signal is visible as a peak at around 2\TeV.

\begin{figure}[tbhp]
  \centering
  \includegraphics[width=.495\textwidth]{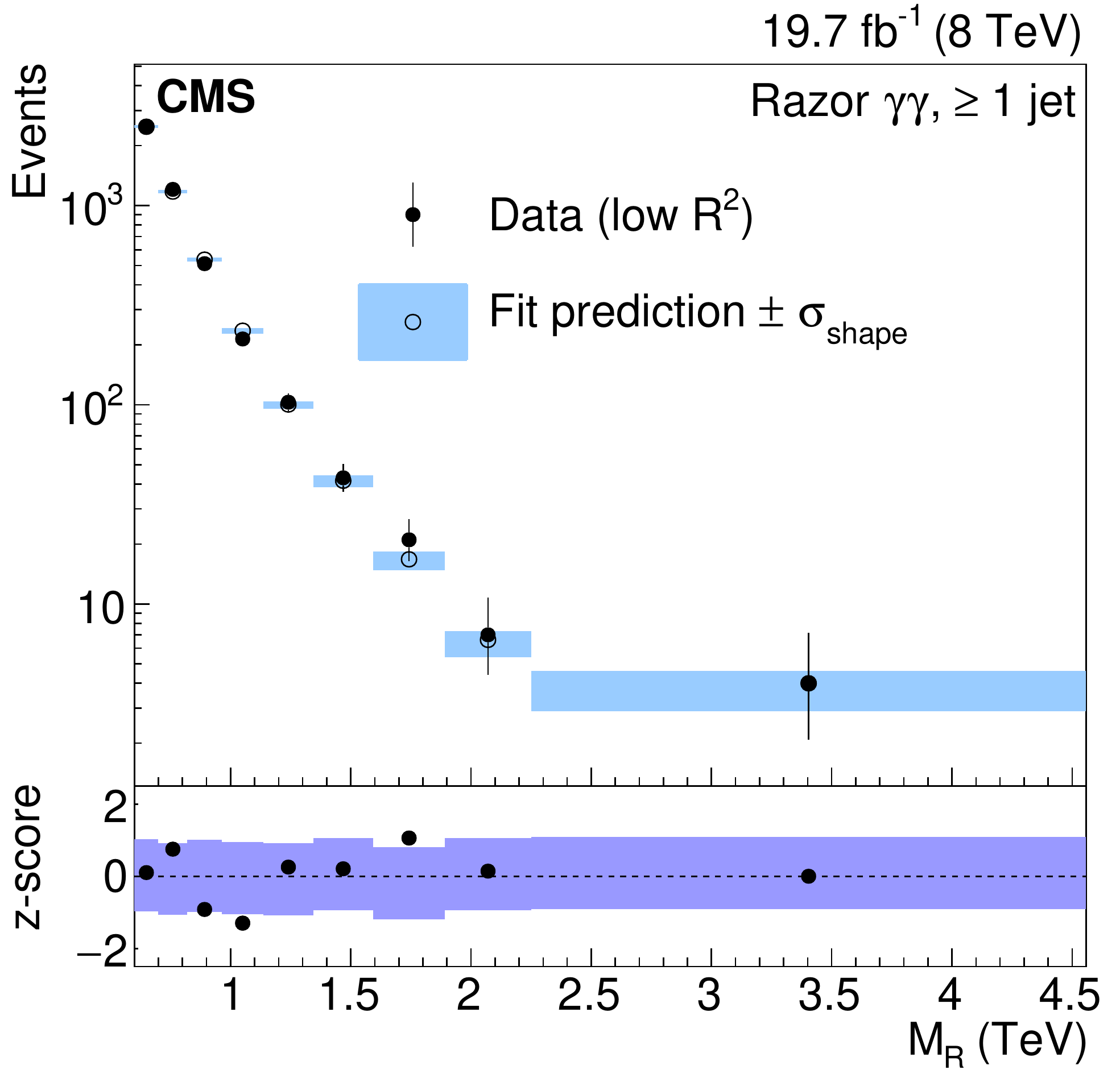}
  \includegraphics[width=.495\textwidth]{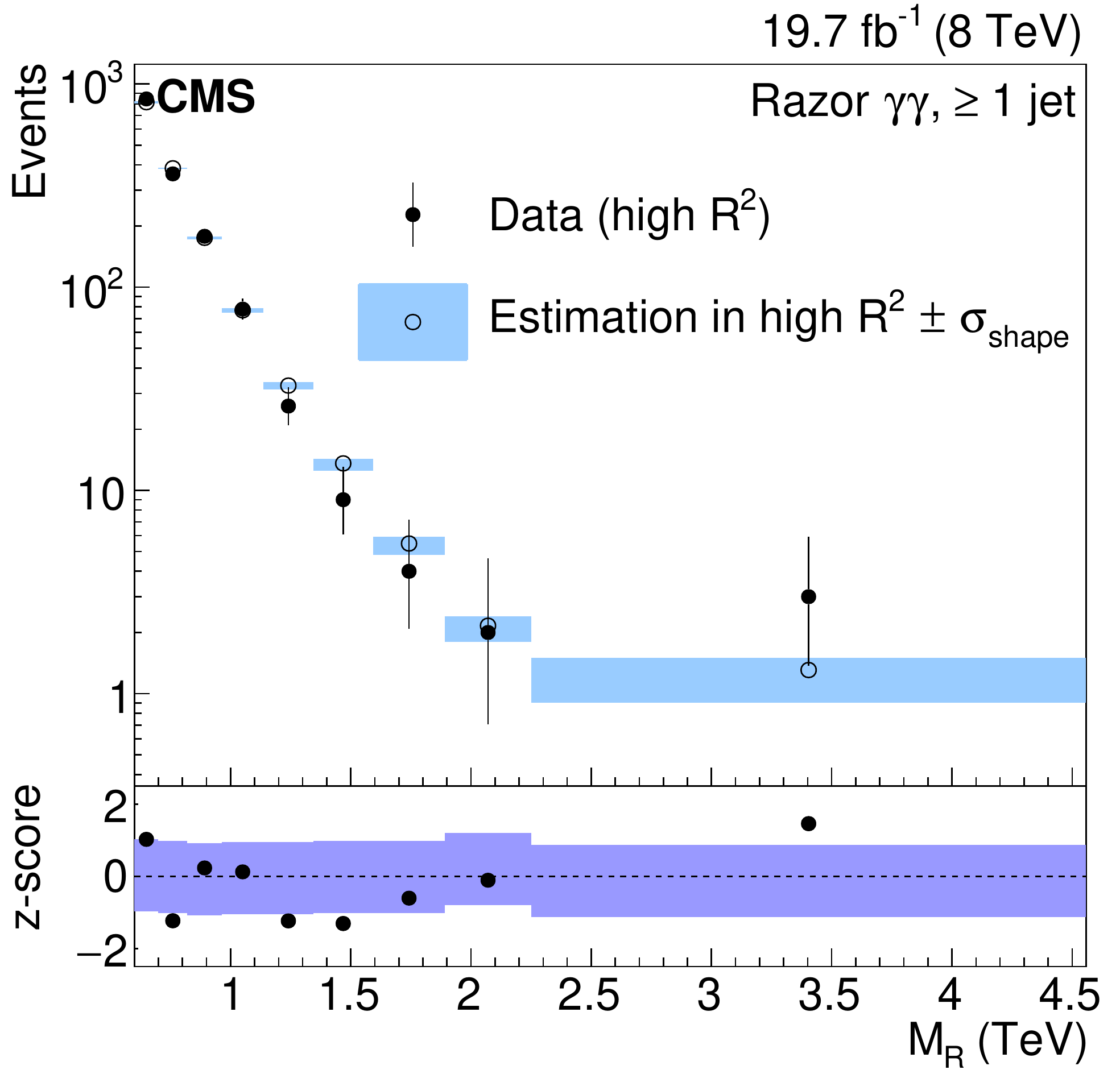}
  \caption{Distribution of $\MR$ for the control
    (\cmsLeft) and signal (\cmsRight) regions.
    The representations of the uncertainty bands, data points, and the information shown in the bottom panel are the same as in
    Fig.~\ref{fig:ffFit}.
    \label{fig:ggFit}
  }
\end{figure}

Figure~\ref{fig:ggFit} (\cmsLeft) shows the result of the fit and the associated
uncertainty band, compared to the data in the control region. The fit
result is then used to derive the background prediction in the signal
region. The comparison of the prediction to the observed data
distribution is shown in Fig.~\ref{fig:ggFit} (\cmsRight). No evidence for a
signal is found. The largest positive and negative deviations from the
predictions are observed for $\MR \gtrsim  2.3$\TeV and $1.1
\lesssim \MR \lesssim 1.9$\TeV, respectively, each
corresponding to a local significance of $\approx$1.5 standard
deviations.

\section{Signal model systematic uncertainties}

Systematic uncertainties in the description of the signals are listed in
Table~\ref{tab:sysSignal}.
Differences between the simulation and data for the photon reconstruction, identification,
and isolation efficiencies are listed as Data/MC photon scale factors.
The uncertainty associated with the parton
distribution functions (PDF) is estimated using the difference in the
acceptance when different sets of PDFs are used~\cite{Alekhin:2011sk,Botje:2011sn,Ball:2012cx,Martin:2009iq,Lai:2010vv}.
 Similarly, different sets of PDFs and different
choices for the renormalization scales yield different predictions for
the expected production cross section.

\begin{table*}[htbp]
\topcaption{The systematic uncertainties associated with
  signal model yields. For the double-photon razor analysis,
  the contributions labeled as ``shape'' have different sizes, depending
  on $\MR$.
  \label{tab:sysSignal}
  }
\centering
\begin{scotch}{l|cc}
Systematic uncertainty & Single photon [\%] & Double photon [\%] \\
\hline
Data/MC photon scale factors  &  1 & 1--2 \\
Trigger efficiency &  2 & 1 \\
Integrated luminosity~\cite{CMS-PAS-LUM-13-001} & 2.6 & 2.6 \\
Jet energy scale corrections~\cite{CMS-JME-10-011} & 2--3 & shape (bin by bin) 2--5\\
Initial-state radiation & 3--5 & $<$1 \\
Acceptance due to PDF & 1--3  & 1--3   \\
Signal yield due to PDF and scales & 5--20 & 1--50  \\
\end{scotch}
\end{table*}

\section{Interpretation of the results}
\label{sec:results}

\begin{figure*}[!thbp]
  \centering
  \includegraphics[width=0.48\textwidth]{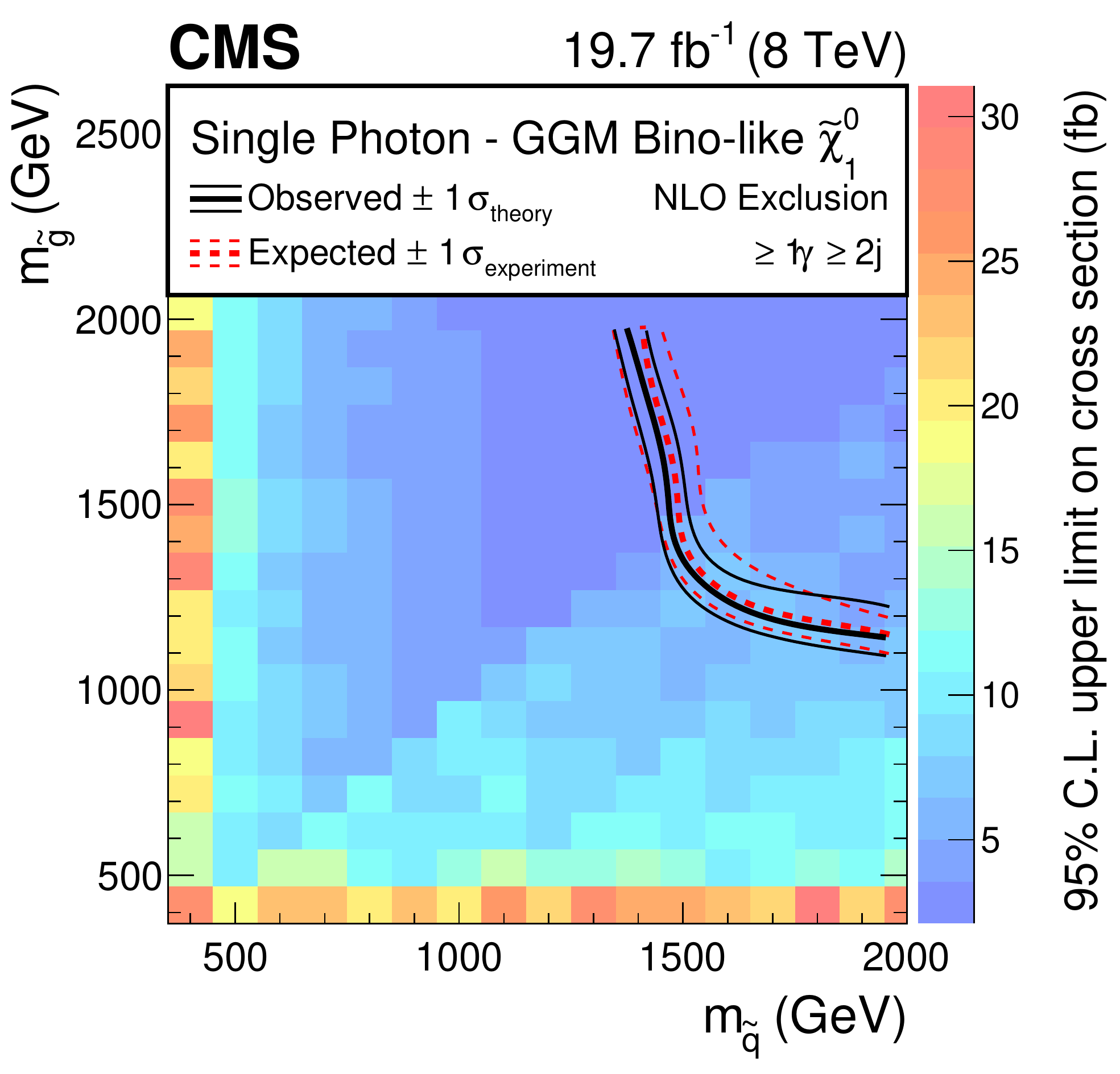}
  \includegraphics[width=0.48\textwidth]{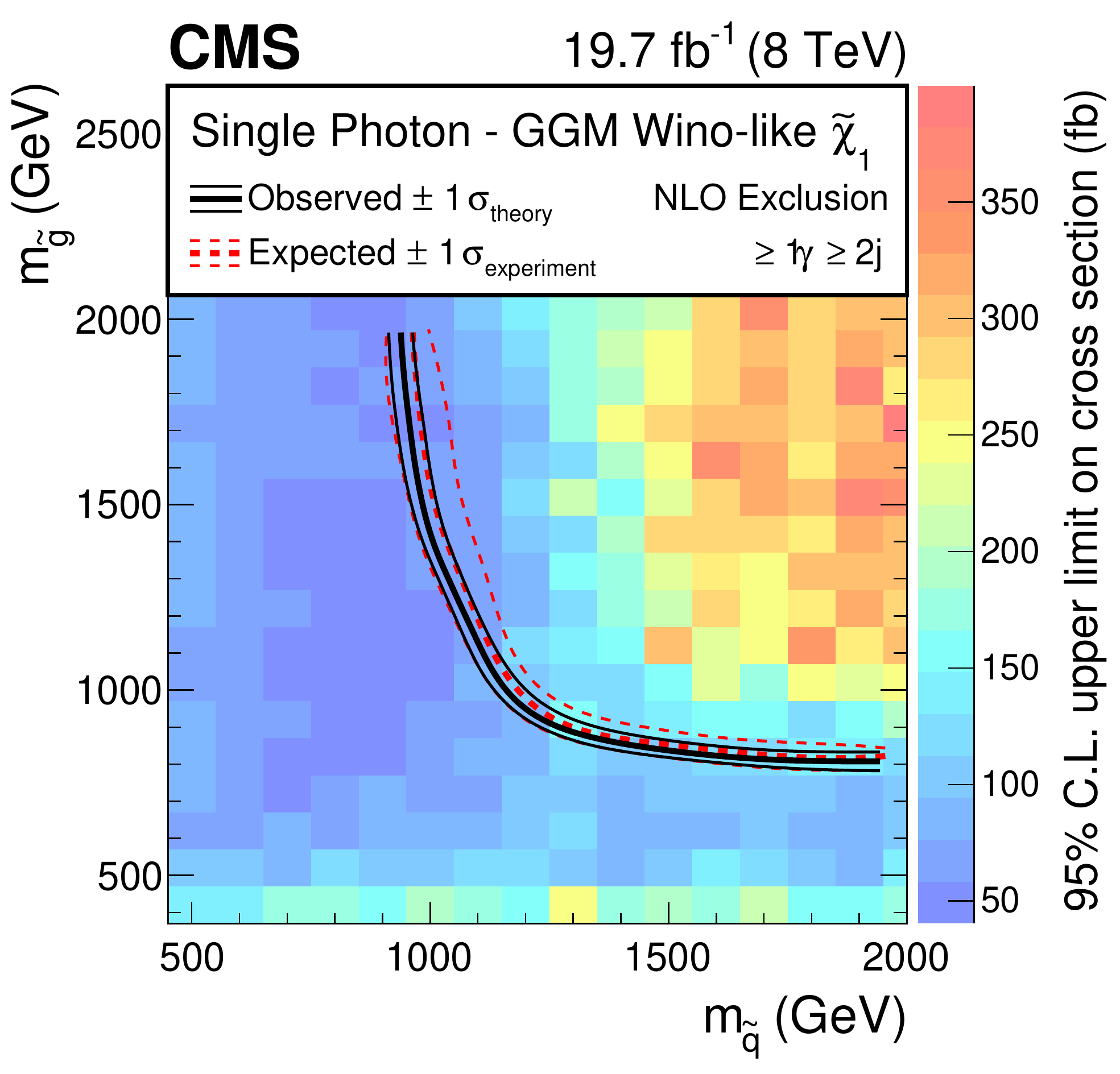}\\
  \includegraphics[width=0.48\textwidth]{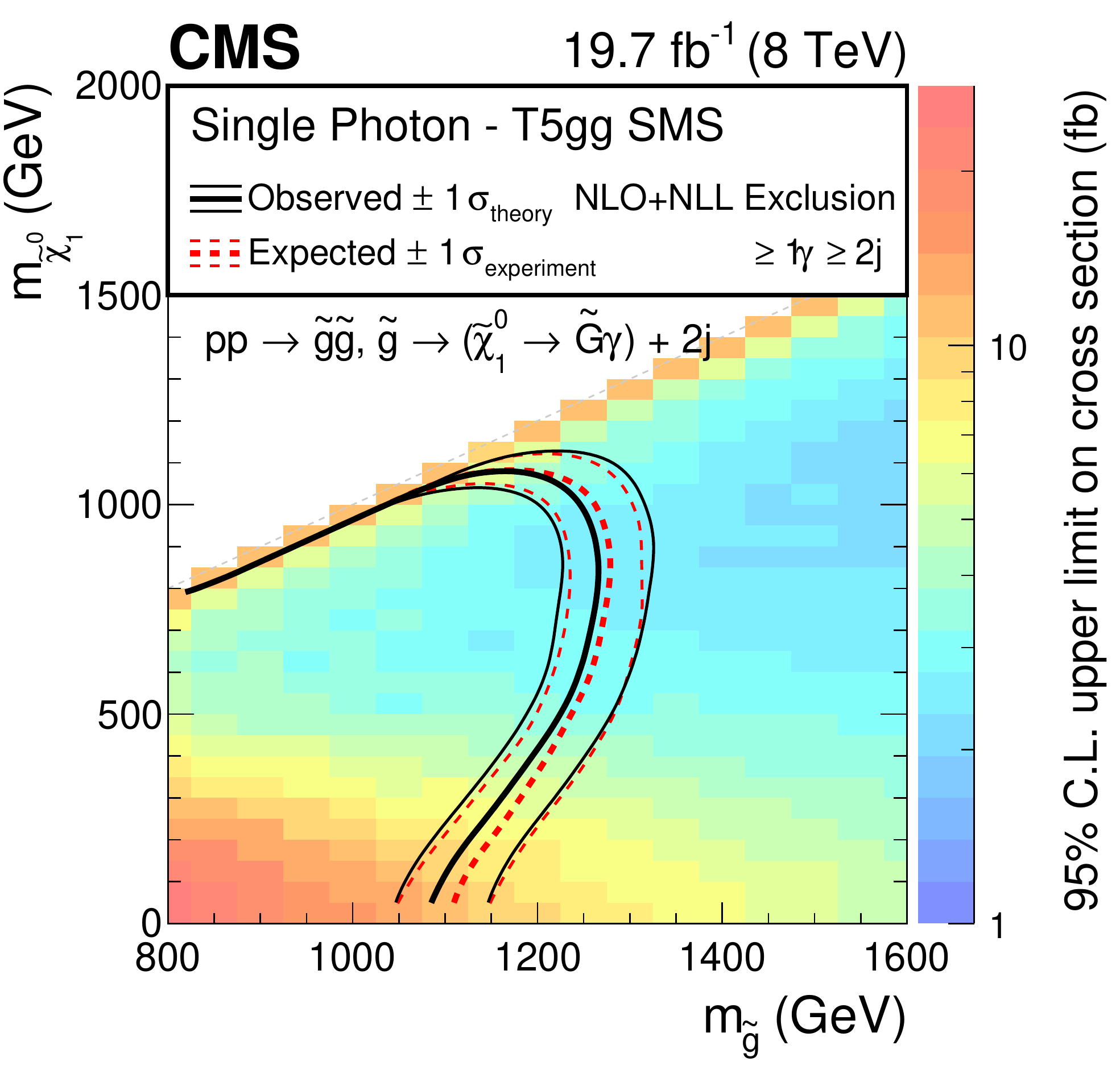}
  \includegraphics[width=0.48\textwidth]{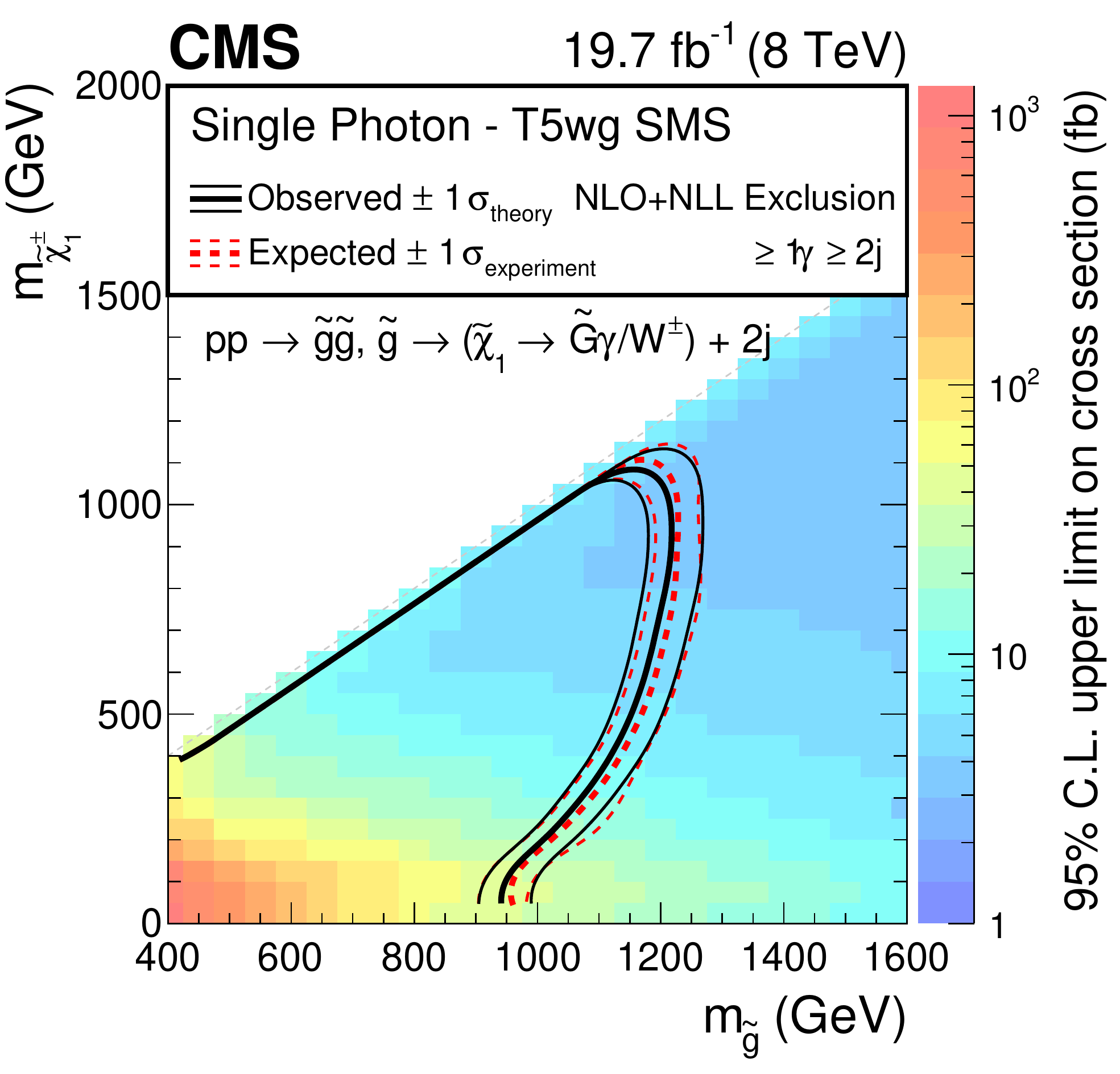}\\
  \caption{
    Single-photon analysis 95\% \CL observed upper limits on the signal cross
    section and exclusion contours in the gluino-squark (top) and gaugino-gluino (bottom) mass plane for the
    GGMbino (top left), GGMwino (top right), T5gg (bottom left), and T5wg (bottom right) models.
    The thick red dashed (black solid) line shows the
    expected (observed) limit. The thin dashed line and band show the 68\%
    \CL range about the expected limit. The solid line quantifies the impact
    of the theoretical uncertainty in the cross section on the observed
    limit. The color scale shows the excluded cross
    section for each set of mass values.
    \label{fig:limitsSingleGamma}
  }
\end{figure*}

The result of the single-photon analysis is used to extract a
limit on the production cross sections of
the GGM and SMS models.
Comparing the excluded cross section
to the corresponding predicted value, a mass limit is derived in the
squark versus gluino mass plane.
This procedure allows comparisons with previous
results~\cite{Chatrchyan:2012bba}.
In the SMS, the limits are derived in the gluino versus gaugino mass plane.
The resulting cross section upper limits and the corresponding exclusion contours
are shown in Fig.~\ref{fig:limitsSingleGamma}.

\begin{figure*}[!thb]
  \centering
  \includegraphics[width=.495\textwidth]{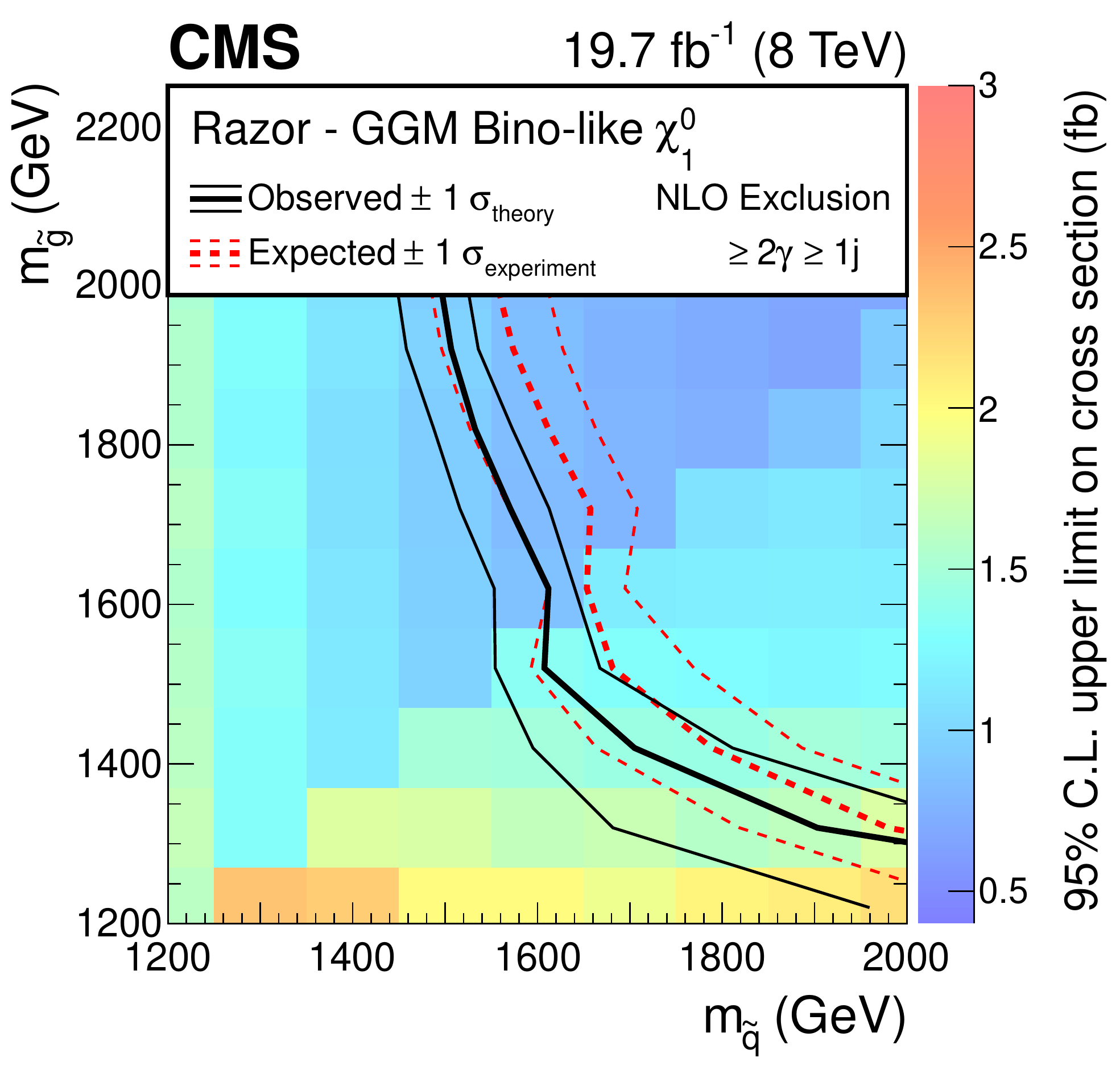}
  \includegraphics[width=.495\textwidth]{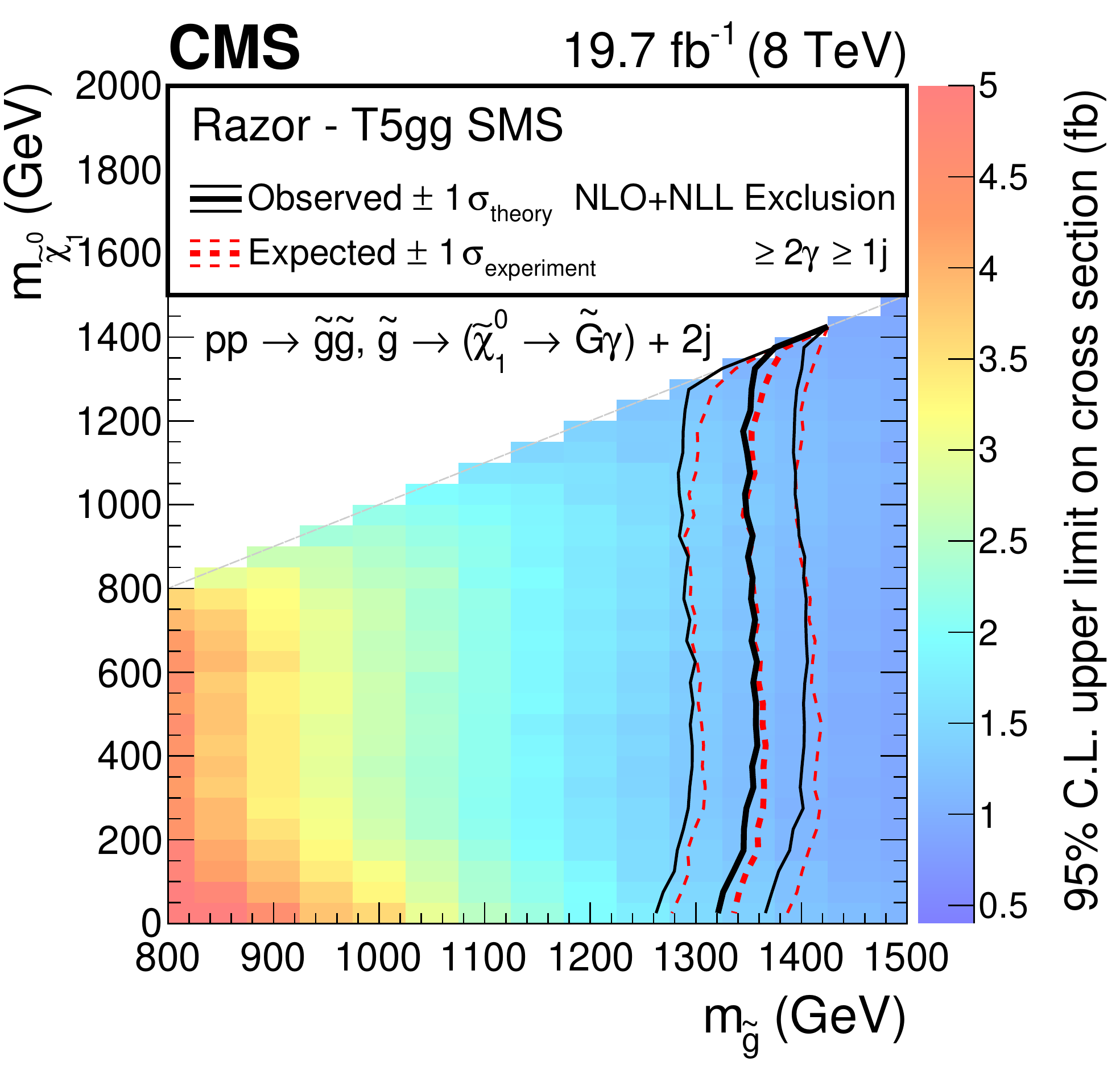} \\
  \caption{Double-photon analysis 95\% \CL observed cross section upper limits
    and excluded mass regions for the GGMbino (left) and T5gg (right) models.
    The representations of excluded regions and cross sections are the
    same as in Fig.~\ref{fig:limitsSingleGamma}.
    \label{fig:limitsRazor}
  }
\end{figure*}

Figure~\ref{fig:limitsRazor} shows the excluded mass regions and the cross
section upper limits for the GGMbino and T5gg models obtained from the
double-photon analysis.

The single- and double-photon analyses are complementary with
respect to the event selection and the search strategy.
While the former is a multichannel counting experiment based on the absolute
prediction of the standard model backgrounds, the latter uses kinematic
information about the razor variables to perform a shape analysis.
The best individual sensitivity is in the wino- and the bino-like neutralino
mixing scenario, respectively. The double-photon analysis performs slightly better compared to the single-photon search in the bino scenario, because of the high-\HT trigger requirement in the single-photon selection.

\section{Summary}
\label{sec:summary}

Two searches for gauge-mediated supersymmetry are presented:
a search based on events with at least one photon
and at least two jets, and a search based
on events with at least two photons and at least one jet.
The single-photon search characterizes a
potential signal as an excess in the tail of the \ETm spectrum beyond
100\GeV, while the double-photon search exploits the razor variables
$\MR$ and $\Rtwo$.
These searches are based on {$\Pp\Pp$} collision data collected with the CMS
experiment at a center-of-mass
energy of $\sqrt{s}=8\TeV$, corresponding to an integrated luminosity
of \lumiData.
No evidence for supersymmetry production is found, and 95\% \CL
upper limits are set on the production cross sections, in the context of
simplified models of gauge-mediated supersymmetry breaking and general gauge-mediation (GGM) models.
Lower limits from the double-photon razor analysis range beyond 1.3\TeV for the gluino
mass and beyond 1.5\TeV for the squark mass for bino-like neutralino mixings in
the studied GGM phase space, extending previous limits~\cite{Chatrchyan:2012bba}
by up to 300 and 500\GeV, respectively.
The limits from the single-photon analysis for wino-like neutralino mixings range beyond
0.8\TeV for the gluino mass and 1\TeV for the squark mass in the same GGM phase space,
extending previous limits by about 100 and 200\GeV.
Within the discussed supersymmetry scenarios, these results represent the current
most stringent limits.

\begin{acknowledgments}

We congratulate our colleagues in the CERN accelerator departments for the excellent performance of the LHC and thank the technical and administrative staffs at CERN and at other CMS institutes for their contributions to the success of the CMS effort. In addition, we gratefully acknowledge the computing centers and personnel of the Worldwide LHC Computing Grid for delivering so effectively the computing infrastructure essential to our analyses. Finally, we acknowledge the enduring support for the construction and operation of the LHC and the CMS detector provided by the following funding agencies: BMWFW and FWF (Austria); FNRS and FWO (Belgium); CNPq, CAPES, FAPERJ, and FAPESP (Brazil); MES (Bulgaria); CERN; CAS, MoST, and NSFC (China); COLCIENCIAS (Colombia); MSES and CSF (Croatia); RPF (Cyprus); MoER, ERC IUT and ERDF (Estonia); Academy of Finland, MEC, and HIP (Finland); CEA and CNRS/IN2P3 (France); BMBF, DFG, and HGF (Germany); GSRT (Greece); OTKA and NIH (Hungary); DAE and DST (India); IPM (Iran); SFI (Ireland); INFN (Italy); MSIP and NRF (Republic of Korea); LAS (Lithuania); MOE and UM (Malaysia); CINVESTAV, CONACYT, SEP, and UASLP-FAI (Mexico); MBIE (New Zealand); PAEC (Pakistan); MSHE and NSC (Poland); FCT (Portugal); JINR (Dubna); MON, RosAtom, RAS and RFBR (Russia); MESTD (Serbia); SEIDI and CPAN (Spain); Swiss Funding Agencies (Switzerland); MST (Taipei); ThEPCenter, IPST, STAR and NSTDA (Thailand); TUBITAK and TAEK (Turkey); NASU and SFFR (Ukraine); STFC (United Kingdom); DOE and NSF (USA).

Individuals have received support from the Marie-Curie program and the European Research Council and EPLANET (European Union); the Leventis Foundation; the A. P. Sloan Foundation; the Alexander von Humboldt Foundation; the Belgian Federal Science Policy Office; the Fonds pour la Formation \`a la Recherche dans l'Industrie et dans l'Agriculture (FRIA-Belgium); the Agentschap voor Innovatie door Wetenschap en Technologie (IWT-Belgium); the Ministry of Education, Youth and Sports (MEYS) of the Czech Republic; the Council of Science and Industrial Research, India; the HOMING PLUS program of the Foundation for Polish Science, cofinanced from European Union, Regional Development Fund; the Compagnia di San Paolo (Torino); the Consorzio per la Fisica (Trieste); MIUR project 20108T4XTM (Italy); the Thalis and Aristeia programs cofinanced by EU-ESF and the Greek NSRF; the National Priorities Research Program by Qatar National Research Fund; the Rachadapisek Sompot Fund for Postdoctoral Fellowship, Chulalongkorn University (Thailand); and the Welch Foundation.
\end{acknowledgments}
\bibliography{auto_generated}

\cleardoublepage \appendix\section{The CMS Collaboration \label{app:collab}}\begin{sloppypar}\hyphenpenalty=5000\widowpenalty=500\clubpenalty=5000\textbf{Yerevan Physics Institute,  Yerevan,  Armenia}\\*[0pt]
V.~Khachatryan, A.M.~Sirunyan, A.~Tumasyan
\vskip\cmsinstskip
\textbf{Institut f\"{u}r Hochenergiephysik der OeAW,  Wien,  Austria}\\*[0pt]
W.~Adam, E.~Asilar, T.~Bergauer, J.~Brandstetter, E.~Brondolin, M.~Dragicevic, J.~Er\"{o}, M.~Flechl, M.~Friedl, R.~Fr\"{u}hwirth\cmsAuthorMark{1}, V.M.~Ghete, C.~Hartl, N.~H\"{o}rmann, J.~Hrubec, M.~Jeitler\cmsAuthorMark{1}, V.~Kn\"{u}nz, A.~K\"{o}nig, M.~Krammer\cmsAuthorMark{1}, I.~Kr\"{a}tschmer, D.~Liko, T.~Matsushita, I.~Mikulec, D.~Rabady\cmsAuthorMark{2}, B.~Rahbaran, H.~Rohringer, J.~Schieck\cmsAuthorMark{1}, R.~Sch\"{o}fbeck, J.~Strauss, W.~Treberer-Treberspurg, W.~Waltenberger, C.-E.~Wulz\cmsAuthorMark{1}
% ...
\vskip\cmsinstskip
% ...
1:~~Also at Vienna University of Technology, Vienna, Austria\\
2:~~Also at CERN, European Organization for Nuclear Research, Geneva, Switzerland\\
% ...

\end{sloppypar}
\end{document}